\documentclass[10pt,aip,jmp,floatfix,preprintnumbers,amsmath,amssymb,amsfonts]{revtex4}

\usepackage{graphicx}
\usepackage{dcolumn}
\usepackage{bm}

\def\C{{\bf C}}
\def\I{{\bf I}}

\def\Q{{\bf Q}}
\def\P{{\bf P}}

\def\Ls{{\cal{L}}}

\def\A{{\bf A}}
\def\F{{\bf F}}

\def\M{{\bf M}}

\renewcommand{\L}{\mathbf{L}}

\usepackage[a4paper,CJKbookmarks,bookmarks=true,bookmarksopen=true]{hyperref}
\hypersetup{
    a4paper,
    pdftitle={Stochastic Structural Stability Theory applied to roll/streak formation in boundary layer shear flow},
    pdfauthor={Brian F. Farrell and Petros J. Ioannou},
    pdfkeywords={},
    bookmarksnumbered,
    pagebackref=true,
    breaklinks=true,
    urlcolor=blue,
    bookmarks=true,
    colorlinks=true,
    linkcolor=red,
    citecolor=blue,          
        }

\begin{document}

\preprint{AIP/123-QED}

\title{Stochastic Structural Stability Theory applied to roll/streak formation in boundary layer shear flow}
\author{Brian F. Farrell} \email{farrell@seas.harvard.edu}
\affiliation{%
Department of Earth and Planetary Sciences, Harvard University\\
24 Oxford Street, Cambridge, MA 02138, U.S.A.}

\author{Petros J. Ioannou}
\email{pjioannou@phys.uoa.gr}
\affiliation{
Department of Physics, University of Athens\\
Panepistimiopolis, Zografos 15784, Greece
}%

\date{\today}
\begin{abstract}
Stochastic Structural Stability Theory (SSST) provides an autonomous, deterministic, nonlinear dynamical system for evolving the  statistical mean state of a turbulent  system.  In this work SSST is applied to the problem of understanding the formation of the roll/streak structures that arise from free-stream turbulence (FST) and are associated with bypass transition in boundary layers.  Roll structures in the cross-stream/spanwise plane and associated streamwise streaks are shown to arise  as a linear instability of interaction between the  FST and the mean flow.    In this interaction incoherent  Reynolds stresses  arising from FST are organized by perturbation streamwise streaks to coherently force perturbation rolls giving rise to an amplification of the  streamwise streak perturbation and through this feedback  to an instability of the combined roll/streak/turbulence complex.  The dominant turbulent perturbation structures involved in supporting the roll/streak/turbulence complex instability are non-normal optimal perturbations with the form of oblique waves.  The cooperative linear instability giving rise to the roll/streak structure arises at a bifurcation in the parameter of STM excitation parameter.  This structural instability eventually equilibrates nonlinearly at finite amplitude and although  the resulting statistical equilibrium streamwise streaks are inflectional the associated flows are stable.    Formation and equilibration of the roll/streak structure by this mechanism can be traced to the non-normality which underlies interaction between perturbations and mean flows in modally stable systems.  
\end{abstract}

\maketitle

\section{Introduction}

The physical mechanism of turbulence in shear flows is not yet comprehensively understood despite  many recent advances in experiment, simulation and theory.  The problem of shear flow turbulence can be divided into two components: transition from the laminar to the turbulent state and maintenance of the turbulent state.  The transition problem results from the lack of an inflection in the velocity profiles of most boundary layer flows.   Inflections are associated by the Rayleigh theorem with existence  of  robust instabilities that continue in viscous flows from the inflectional instability of the same velocity profile in an inviscid flow.  The problem of observed robust disturbance growth in perturbation stable shear flows was solved when it was recognized that the non-normality of the underlying linear dynamics of shear flows allows perturbation growth in the absence of exponential instability.  The concept of transient growth in shear flow has roots in the classical work of Kelvin and Orr \citep{Kelvin-1887b, Orr-1907}  who used analytical solutions of perturbation dynamics in idealized shear flows to provide example solutions demonstrating transient growth.  This early work remained obscure presumably due to lack of a convincing physical application. Transient growth concepts were first applied in the modern context of linear operator non-normality to understanding the three dimensional baroclinic turbulence in the  midlatitude atmospheric jet after comprehensive data collected beginning in the middle of the last century for the purpose of weather forecast provided convincing evidence that  turbulence in the midlatitude jet stream was maintained by growth processes unrelated to exponential instability \citep{Farrell-1982, Farrell-1985, Farrell-1989}.   In the context of laboratory shear flows although three dimensional perturbations in the form of a roll/streak structure were observed in boundary layers (Townsend, 1956; Kline et al., 1967; Blackwelder and Eckelmann, 1979; Robinson, 1991) and related to the nonmodal lift-up growth mechanism \citep{Landahl-1980}  similarly comprehensive observational evidence for the mechanism of nonmodal growth in boundary layer flows awaited the advent of direct numerical simulation (DNS) at Reynolds numbers O(1000), for which turbulence is maintained in shear flow, and particle image velocimetry (PIV) of turbulent laboratory shear flows.   The methods of non-normal operator analysis and optimal perturbation theory were first applied  in the context of laboratory shear flows to two dimensional disturbances \citep{Farrell-1988a}.  It was believed at the time  that secondary instability of finite amplitude two dimensional equilibria were the mechanism of transition \citep{Pierrehumbert-1986, Bayly-etal-1988} and it was  shown that these unstable two dimensional nonlinear finite amplitude equilibria could be readily excited by even very small optimal initial perturbations\citep{Butler-Farrell-1994}.  However, it became increasingly apparent from observation and simulation that the finite amplitude structures associated with transition are three dimensional  and analysis of three dimensional optimal perturbation growth followed \citep{Butler-Farrell-1992, Butler-Farrell-1993, Farrell-Ioannou-1993a, Farrell-Ioannou-1993b,Reddy-Henningson-1993,Trefethen-etal-1993,Schmid-Henningson-2001}. These analyses revealed that the optimally growing three dimensional structure is associated with cross-stream/spanwise rolls and associated streamwise streaks and is related to the linear lift up mechanism.  The remarkable convected coordinate  solutions for perturbation growth in unbounded shear flow \citep{Kelvin-1887b} allow closed form solution for the scale independent structures producing optimal growth in three dimensional shear flow \citep{Farrell-Ioannou-1993a, Farrell-Ioannou-1993b}.  These closed form optimal solutions in unbounded shear flow confirm the result found numerically in bounded shear flows that for sufficiently long optimizing times streamwise rolls produce optimal energy growth while for short optimizing times the optimal perturbations  are oblique wave structures that  synergistically exploit both the two dimensional shear and the three dimensional lift up mechanisms producing  vortex cores oriented at an  angle of approximately $60$ degrees from the spanwise direction. And indeed, the roll/streak and oblique accompanying structure complex that is predicted to produce optimal growth by analysis of non-normal perturbation dynamics of shear flows has been convincingly seen in both observations and simulations 
\citep{Klebanoff-etal-1962, Schoppa-Hussain-2000, Adrian-2007, Wu-Moin-2009} and shown to be essentially related to the  non-normality of shear flow dynamics
\citep{Kim-Lim-2000, Schoppa-Hussain-2002}.

Although the mechanism of non-normal growth has been clarified, and its importance in bypass transition and maintenance of the turbulent state is widely if not universally accepted, the route by which non-normality leads to formation of the roll/streak structure and the part played by this coherent structure, its nonlinear equilibration and its stability, in the transition to turbulence and maintenance of the turbulent state remains to be determined.

The most direct mechanism exploiting non-normality to form roll/streak structures is introduction of an optimal perturbation into the flow, perhaps  by using a  trip or other device \citep{Butler-Farrell-1992, Reddy-Henningson-1993, Trefethen-etal-1993}.  A related approach is to stochastically force the flow, with stochastic forcing regarded as modeling surface roughness or  free stream turbulence (FST).  This mechanism can be analyzed
using stochastic turbulence modeling  (STM) \citep{Farrell-Ioannou-1993e, Farrell-Ioannou-1993-unbd, Farrell-Ioannou-1994b, Farrell-Ioannou-1998a, Farrell-Ioannou-1998b, Bamieh-Dahleh-2001, Jovanovic-Bamieh-2005, Hoepffner-Brandt-2008}.  Because of the non-normal nature of perturbation growth in shear flow,  stochastic turbulence models are  closely related to optimal perturbation dynamics.    In conventional stochastic turbulence models the roll/streak structure is envisioned to arise from chance occurrence of optimal or near optimal perturbations in the stochastic forcing\citep{Bamieh-Dahleh-2001,Hwang-Cossu-2010, McKeon-etal-2010, Hwang-Cossu-2010a}.   These mechanisms exploit the linear non-normal growth process directly.   However, the ubiquity of streak formation suggests, as argued by Shoppa and Hussein \citep{Schoppa-Hussain-2000},  that some form of instability process underlies the formation of streaks,  that this instability involves an intrinsic association between the  roll/streak structure and  associated oblique waves and vortices, and that this three dimensional instability must differ qualitatively from the familiar laminar shear flow instability.   Indeed,  from a comparison of experiment with simulation Anderson et al. \citep{Andersson-etal-1999} conclude that  the evidence ``..corresponds to some fundamental mode triggered in the flat-plate boundary
layer when subjected to high enough levels of free-stream
turbulence..".  Previously proposed exponential instability mechanisms include centrifugal instability \citep{Brown-Thomas-1977} and the Craik-Leibovich instability 
\citep{Phillips-etal-1996}.
Proposed algebraic growth mechanisms involve a streamwise average torque produced by  interaction of discrete oblique waves\citep{Benney-1960,  Jang-etal-1986}.

All these streak growth mechanisms rely on the dominant non-normal process in shear flow which is lift up of mean streamwise velocity by perturbation cross-stream velocity, while differing in the manner in which this cross-stream velocity arises.     

The cross-stream/spanwise roll structure provides a powerful mechanism for forming streamwise streaks in shear flows whether episodically forced by an initial condition or continuously forced by an oblique wave structure.  However, in the absence of feedback from the streak back to the roll this powerful streak formation mechanism does not result in instability although because of the large streak growth produced by a cross-stream/spanwise roll perturbation, placing even a very weak coupling of the streak to the roll, such as by a small spanwise frame rotation, results in destabilization \citep{Komminaho-etal-1996, Farrell-Ioannou-2008-baroclinic}.  Turbulent Reynolds stress provides an alternative  mechanism for producing the feedback between the streak and roll needed to destabilize the roll/streak structure.  Indeed, if we observe a turbulent shear flow in the
cross-stream/spanwise plane at a fixed streamwise location we see that at any instant  there is a substantial torque from Reynolds stress divergence forcing cross-stream/spanwise rolls.  The problem is that this torque is not systematic and so it vanishes in temporal or streamwise average.  However, in the presence of a perturbation streak the symmetry in the spanwise direction is broken and the torque from Reynolds stress divergence  can become organized to produce the positive feedback between the streak and roll required to destabilize this structure by continuously and coherently exploiting the powerful non-normal roll/streak amplification mechanism.  The existence of this mechanism for destabilizing the roll/streak structure in turbulence makes it likely that some dynamical perturbation complex exists to exploit it.   In this work we prove by construction that this is so by deriving a system of equations   eigenanalysis of which reveals the unstable roll/streak/turbulence structure that is responsible for destabilizing turbulent shear flow to streak formation.  This emergent instability can be understood as a synthesis of the streak formation mechanisms described above in which FST, rather than itself constituting the cross-stream velocity linearly forcing streak growth, instead is organized by the  perturbation streak into oblique waves that quadratically force the cross-stream/spanwise roll by inducing a Reynolds stress torque linearly proportional to streak amplitude thereby producing  an emergent exponential instability of the combined roll/streak/turbulence complex.  As a boundary layer develops in the presence of FST this exponential instability of organization of the roll/streak complex is the first non-viscous instability to occur.   We remark that this coupling  in the highly non-normal boundary layer shear flow of the roll and streak components by turbulent stresses resulting in destabilization of the roll/streak structure is analogous to the coupling of the torroidal and poloidal components of the magnetic field by the turbulent $\alpha$ effect which destabilizes the induction equation  in the magnetic dynamo problem\citep{Parker-1955}.

The challenge is to find a method of stability analysis  analogous to the method of modes for laminar flow instability, but applicable to the emergent turbulence/mean flow interaction instability.   
Specifically, analysis of the cooperative instability of the roll/streak/turbulence complex  requires constructing a dynamical system for evolving the consistent statistical mean of a turbulent state.  What is required is a  physically correct while at the same time a computationally accessible approximation to the evolution of the trajectory of the  probability density function (pdf) of a turbulent system in phase space.  We refer to this dynamical system as the stochastic structural stability theory (SSST) system.  The approximate pdf  of the turbulent state, which is carried forward in time by the SSST system dynamics, often evolves to a fixed point corresponding to a statistically steady turbulent state.   
This method for analyzing the dynamics of the pdf of a turbulent system was developed to study the phenomenon of spontaneous jet formation at global scale in planetary atmospheres \citep{ Farrell-Ioannou-2003-structural, Farrell-Ioannou-2007-structure, Farrell-Ioannou-2008-baroclinic} and has also been applied to the problem of spontaneous jet formation from drift wave turbulence in magnetic fusion devices\citep{Farrell-Ioannou-2009-plasmas}.  In SSST the turbulence is simulated using a STM in which intrinsic excitation of perturbations by nonlinear scattering and extrinsic excitation by FST are parameterized as stochastic \citep{Farrell-Ioannou-1993d, Farrell-Ioannou-1996a, DelSole-Farrell-1996, DelSole-04}.  The STM provides an evolution equation for the quadratic statistics of the turbulent eddy field associated with a mean flow.   In the STM the eddy field is expressed  in terms of a covariance matrix from which the Gaussian probability density function approximation for the turbulence variance and quadratic fluxes can be obtained.   Coupling a time dependent STM to an evolution equation for the streamwise mean  roll/streak/shear complex produces a nonlinear dynamical system for the co-evolution of the roll/streak/shear and the self-consistent quadratic statistics of its associated turbulence: this is the SSST system.    

The SSST equations incorporate a stochastic turbulence model but these equations are themselves deterministic and autonomous with dependent variables the streamwise mean roll/streak/shear complex and the  streamwise mean covariance of the turbulence.  The perspective on shear flow stability provided by these equations differs from the more familiar perspective based on perturbation stability of stationary laminar flows.  In fact,  the primary perturbation instability in SSST has no counterpart in the stability theory of laminar flow; it is rather a cooperative instability  in which the evolving roll/streak/shear complex organizes the background turbulence covariance to produce flux divergences configured to amplify the roll/streak/shear complex leading to an emergent coupled roll/streak/shear plus turbulence instability that does not involve perturbation instability of the streak.  The SSST equations approximate the nonlinear  streamwise mean dynamics  of the coupled roll/streak/shear plus turbulence complex and this system in many cases supports equilibration of the emergent roll/streak/shear complex and its consistent turbulence field at finite amplitude.  Equilibrium between a mean flow and its associated field of  turbulence requires that the momentum flux divergence arising from the turbulent and mean velocities produce a stationary state of balance  with the streamwise mean flow forcing and dissipation.   The remarkable fact is that the turbulence, which depends on the roll/streak/shear complex, and the roll/streak/shear complex, which depends on the turbulence, quite generally adjust to produce such balanced states.

While the solution trajectory of an initially unstable SSST state generally converges to a fixed point representing a state of balance among the mean flow forcing and advection, the turbulent Reynolds stress divergence, and the damping; these finite amplitude equilibria may lose structural stability as a function of  the STM excitation parameter and this instability then leads either to another equilibrium or to a time dependent limit cycle or chaotic solution \citep{Farrell-Ioannou-2003-structural, Farrell-Ioannou-2007-structure, Farrell-Ioannou-2008-baroclinic, Farrell-Ioannou-2009-equatorial, Farrell-Ioannou-2009-closure}.  Chaotic trajectories of the SSST system correspond not to the familiar chaos of an individual turbulent state trajectory but rather to chaos of the ensemble mean turbulent state.  
An example of this kind of chaos is the irregular fluctuation of the mean flow/turbulence complex seen in  drift wave turbulence \citep{Mazzucato:1996}.    

In this work we concentrate on the emergence of roll/streak structures in boundary layers as an instability of interaction between FST and the streamwise mean flow and on the mechanism by which this instability equilibrates at finite amplitude to maintain a statistically stable mean roll/streak/turbulence structure.   Stable roll/streak structures are initiated as an instability at a minimum intensity of FST and are maintained as statistically stationary structures over an interval of STM excitation parameter.  The time dependent state initiated at higher levels of FST will be reported on elsewhere.

\section{Review of streamwise vortex forcing by oblique waves}

The emergent SSST streak instability mechanism can be understood as a synthesis of mechanisms previously investigated in the study of streak formation in boundary layers  and we begin by reviewing  the mechanism of streamwise vortex forcing by oblique waves which  has been used\citep{Benney-1960, Jang-etal-1986, Schmid-Henningson-1992}   to explain the streamwise mean vortex circulations observed by Klebanoff et al.  \citep{Klebanoff-etal-1962}.  These works ascribe streak formation to interaction of a specific mixture of  Tollmien-Schlichting (TS) waves  and oblique perturbations chosen to produce through quadratic advective interaction a zero streamwise wavenumber  component.  Consider a fluctuating eddy field and its associated Reynolds stresses  $\overline{{u}_i {u}_j}$, where ${u}_i$ is one of the three components of the  eddy velocity field, $u$ in the streamwise, $x$, direction; $v$ in the cross-stream, $y$, direction; and $w$ in the spanwise,  $z$,  direction.  Here and in the sequel an overline denotes the streamwise average.  Divergence of the Reynolds stress induces a streamwise mean force:

\[
F_i = - \frac{\partial  \rho \overline{{u}_i{u}_j}}{\partial x_j}.
\]

The mean spanwise force is
\[
F_z= -\frac{\partial \rho \overline{v w}}{\partial y}-\frac{\partial \rho \overline{w^2}}{\partial z}
\]
and the mean cross-stream force is:
\[
F_y= -\frac{\partial \rho \overline{v w}}{\partial z}-\frac{\partial \rho \overline{v^2}}{\partial y}~.
\]
Consequently the streamwise component of the mean torque  $ \vec{G}= \nabla \times \vec{F}$ is: 

\begin{eqnarray*}
G_x &= & \frac{\partial F_z}{\partial y}-\frac{\partial F_y}{\partial z} \\
&=& \left ( \frac{\partial^2  }{\partial z^2}-\frac{\partial^2   }{\partial y^2}  \right ) \rho \overline{vw}+
\ \frac{\partial^2  }{\partial y \partial z} \rho \left ( \overline{v^2} - \overline{w^2} \right )~.
\label{eq:Gx}
\end{eqnarray*}

If the eddy field is spanwise homogeneous no streamwise mean torque is present.  However, it is a remarkable fact  that quite generally in three dimensional flows when spanwise symmetry is broken  mean torques arise that maintain streamwise mean vortices.  

As an example it is instructive to review the case of an unbounded domain  with  a flow field consisting of two oblique structures in the $x,z$ plane with  wave vector $(k,m)$ inclined at an angle   
$\Theta= \pm \tan^{-1} (m/k)$ to the streamwise direction:

\begin{equation}
u=u_0 \cos mz e^{ i kx + i l y}~,~v=v_0 \sin mz e^{ i kx + i l y}~,~w=w_0 \cos mz e^{ i kx + i l y}~~,
\label{eq:oblique}
\end{equation}

With velocity components satisfying non-divergence:
\[
i k u_0 +i l v_0  +  m w_0 = 0~.
\]

The Reynolds stresses for this flow are:

\[
\overline{vw}= \frac{1}{4} \Re(v_0 w_0^*) \sin 2 m z~~,~~\overline{v^2}= \frac{1}{2} |v_0|^2 \cos^2 m z~~,~~
\overline{w^2}= \frac{1}{2} |w_0|^2 \sin^2 m z~.
\]

The spanwise dependence of the Reynolds stresses implies a streamwise mean force $F_y(z)$ in the cross-stream plane inducing  streamwise   mean  toque:
\[
G_x= - \rho m^2 \Re(v_0 w_0^*) \sin 2 m z~.
\]

Note that this torque has double the spanwise wavenumber of the perturbation field.  There is no streamwise torque induced by a single oblique wave when $m=0$.  Perhaps less obvious is that  no Reynolds stress torques are induced  by streamwise rolls with $k=0$  because continuity requires $v_0$ and $w_0$ be in quadrature and therefore $\Re(v_0 w_0^*)=0$.

The vanishing of the torque in these two limiting cases is made clear by using the continuity equation to write
\[
G_x=- \rho \frac{k m^2}{l} \Re(u_0 w_0^*) \sin 2 m z~,
\] 
it is also immediate from this expression  that for given square wavenumber $k^2+m^2$ the torque is maximized for oblique waves with $\Theta \equiv \tan ^{-1} (m/k) = 54.7 ^0$.  

For unbounded  constant shear  flow, $U= \alpha y$, plane wave solutions in closed form exist \citep{Kelvin-1887b}.  For such a  flow it can be easily shown that the Reynolds stress for an initial eddy field in the form
of the oblique perturbations in the form   (\ref{eq:oblique})  is:
\[
\overline{vw}(t) = \frac{1}{4} \frac{k^2+l^2+m^2+l^2}{k^2+m^2+(l-\alpha k t)^2} e ^{-g (t) } \Re(v_0 w_0^*) \sin 2 m z~,
\]
where $e^{-g (t)}$ is the decay  due to dissipation. The time dependent Reynolds stress produced by an oblique wave in constant shear flow produces torque with constant spatial structure and as a result coherent mean streamwise torque over its lifecycle.   Also it can be shown using this closed form solution \citep{Farrell-Ioannou-1993a} that  energy growth over short time intervals is maximized for oblique perturbations with $\Theta= 63^0$  which is very close to the orientation of the oblique wave maximizing the induced torque.  It is consistent then that near $\Theta= 63^0$ lies the oblique waves that producing the greatest mean streamwise torques when integrated over their life cycle.   

\begin{figure*}
 
\includegraphics[width=7in]{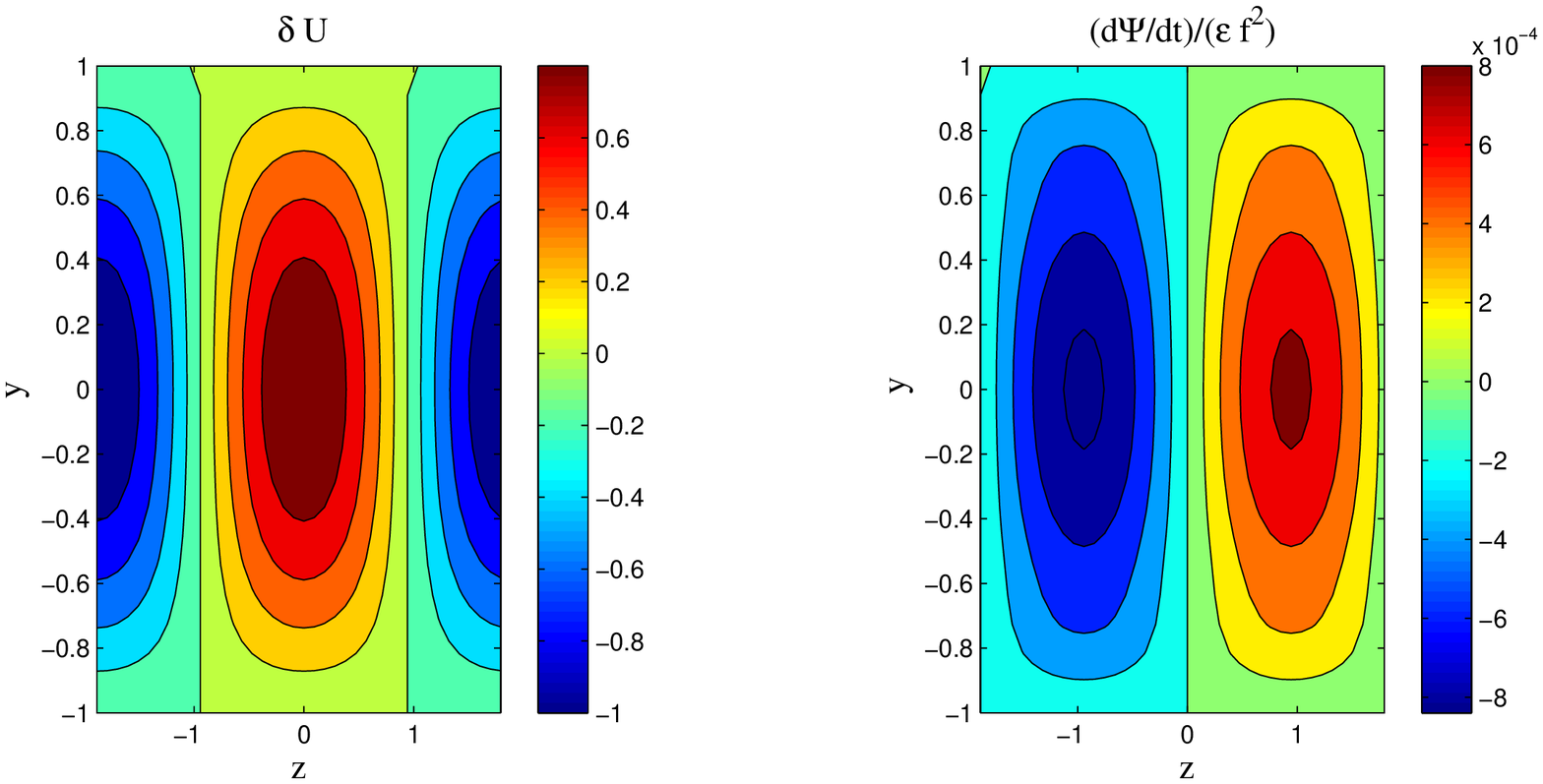}
\vspace*{-1mm} 
\caption{ 
(Color online) Left:  Streamwise flow perturbation 
$\delta U= \epsilon \cos ( \pi y / ( 2 L_y )) \cos (2 \pi z / Lz)$ 
imposed on a background Couette flow to examine the mechanism of
turbulent Reynolds stress organization. Right:  Resulting equilibrium Reynolds stress divergence
induced tendency in the cross-stream/spanwise streamfunction, $d \Psi / dt $,   normalized by the 
mean flow perturbation amplitude,  $\epsilon$, and  the square of the STM excitation parameter parameter,   $f$.
Imposition of a spanwise perturbation  breaks the spanwise symmetry of the Couette flow producing a coherent  streamwise torque proportional to both the mean flow perturbation and to the eddy field variance. The channel dimensions are  $L_y=1$, $L_z=1.2 \pi$, the Reynolds number is $R=400$, and the eddy field is at streamwise wavenumber $k=2 \pi/ (1.75 \pi)$.} 
\label{fig:benney_random}
\end{figure*}

However, the fact that a specific pair of plane oblique waves force streamwise vortices does not explain the presence of streamwise vortices when the mean flow and the perturbation field statistics are spanwise uniform.   Remarkably, while no mean streamwise torque arises from turbulence in a fluid with only a constant shear in the cross-stream direction,  
the slightest spanwise structure in the streamwise flow induces coherent streamwise mean torques. An example of this phenomenon  obtained by imposing the mean streamwise perturbation test function
$\delta U = \epsilon \cos ( \pi y /(2 L_y)) \cos (2 \pi z / L_z)$  on a Couette flow is shown in 
Fig.\ref{fig:benney_random}.  This result was obtained using the STM described in section  III by first calculating the equilibrium perturbation velocity covariance  maintained by stochastic excitation in the perturbed streamwise mean flow
 and  then obtaining  from this perturbation velocity covariance the  resulting Reynolds stress
induced torque. This coherent torque arises because the eddy field is modified by the spanwise variation of the streamwise streak.   This induced torque in turn modifies the mean flow which by interaction with the eddy field produces a modified torque.  Most mean flow perturbations organize torques that do not exactly amplify that mean flow perturbation, as is the case for the perturbation in Fig. \ref{fig:benney_random}.  However,   exponential growth of streamwise mean flows and associated eddy fields would result,  at least for sufficiently small perturbations,  if the mean flow perturbation were to organize precisely the eddy field required for its amplification.    We show constructively below that these  cooperative instability structures arise quite generally by using an eigenanalysis to solve for them.   Boundary layer flows in general support this structural instability in the presence of sufficient FST.  For a range of FST intensities this structural instability equilibrates at finite amplitude producing stable but strongly inflected spanwise streaks.

\section{Formulation of the composite roll/streak/turbulence dynamics}

We decompose the velocity fields  into streamwise mean components  (indicated uppercase) and perturbations (indicated lowercase) so that the total streamwise velocity in the $x$ direction
is  $U(y,z,t) + u(x,y,z,t)$, the cross-stream velocity in the $y$ direction is $V(y,z,t) + v(x,y,z,t)$, and the spanwise velocity in the $z$ direction is $W(y,z,t) + w(x,y,z,t)$.  The flow is confined to the channel $|y| \le 1$,  $|z|\le  L_z/2$ and the Reynolds number, $R$,
is based on the half cross-stream channel distance $L_y=1$. 
With constant velocity  channel walls forcing a Couette flow as
a laminar equilibrium the boundary conditions  are 
$U(\pm 1,z,t) = \pm 1$, $V(\pm 1,z,t)=W(\pm 1,z,t) = 0$ and $u(x,\pm 1,z,t)=v (x,\pm 1,z,t) = w(x,\pm 1,z,t) = 0$ which imply
that the normal vorticity
$\eta= \partial_z u-\partial_x w$  satisfies the boundary condition $\eta(x,\pm 1,z,t)=0$ and from continuity,  $v_y(x,\pm 1,z,t) =0$.  Periodic boundary conditions are imposed in the spanwise direction.

We  write  perturbation equations about the streamwise mean flow, $U(z,y,t)$, neglecting the small spanwise and cross-stream mean flows, as it can be verified that $||V||,||W|| <<||U||$.  With this simplification the perturbation  equations can be reduced, following steps similar to those used in the derivation of the Orr-Sommerfeld and Squire system \citep{Waleffe-1995,Schmid-Henningson-2001} , to two equations in the  normal velocity, $v$, and normal vorticity, $\eta$. The resulting equations for the perturbations in these variables are: 
\begin{subequations}
\begin{eqnarray}
 v_t + \Delta^{-1}\left (  U  \Delta _x + U_{zz} v_x  + 2 U_z v_{xz} - U_{yy} v_x - 2 U_z w_{xy}- 2 U_{yz} w_x  - \Delta \Delta v/R \right)  &=&  F_v~, \label{eq:pv} \\
\left.\eta_t  +  U \eta_x  - U_z v_y + U_{yz} v + U_y v_z+ U_{zz} w   - \Delta \eta/R \right) & = & F_{\eta}~.
\label{eq:peta}
\end{eqnarray}
\end{subequations}
where $\Delta^{-1}$ is the inverse of the  Laplacian,  $\Delta \equiv \partial_{xx}^2+\partial_{yy}^2+\partial_{zz}^2$. 
On the RHS  $F_v$ and $F_{\eta}$ represent 
deviations of the perturbation-perturbation advection   from its streamwise
average,  as is the convention in Reynolds averaging, together with  input  from FST.  
A parameterization of these terms will be specified in the next section.

The  spanwise  and streamwise velocity  expressed in terms of the variables $v$ and $\eta$ are:
\begin{equation}
\Delta_h w= -v_{yz}-\eta_x ~~, ~~\Delta_h u = -v_{yx}+ \eta_z,  \label{eq:wudiag}
\end{equation}
\noindent in which $\Delta_h\equiv \partial_{xx}^2+\partial_{zz}^2$ denotes the horizontal Laplacian.

The mean streamwise  flow, $U$, evolves  according to
\begin{equation}
 U_t = -( U V+ \overline{uv} )_y - ( U W + \overline{uw} )_z + \Delta U/R~. \label{eq:MU}
\end{equation}
The mean  streamwise  flow $U$ is maintained against dissipation by the streamwise component of the force from the eddy Reynolds stresses 
and by the acceleration induced by the  mean roll circulation $-( U V )_y - ( U W)_z$ which can be written equivalently  as
$-U_y V -U_z W$.  In spanwise independent flows this term reduces to $-U_y V$ and represents the familiar lift up  mechanism
\citep{Landahl-1980}.   In order to evolve the streamwise mean flow the eddy Reynolds stresses
and the  fields $V$ and $W$ associated with the roll circulation are required.   We define the streak perturbation as the component of $U$ that deviates  from its spanwise average, $[U]= \int_0^{L_z} U dz / L_z$, so that that the velocity associated with the streak is $U_s = U-[U]$. 

The streamwise mean cross-stream and spanwise velocities;  $V$ and  $W$ respectively, can 
be obtained from the mean streamfunction $\Psi(y,z,t)$:
\begin{equation}
V= - \Psi_z~~, ~~W = \Psi_y~,
\end{equation}
which evolves  according to the  equation for mean  streamwise vorticity $\Delta \Psi$. 
\begin{equation}
\Delta \Psi_t=  (V W + \overline{ vw} )_{zz} - (V W + \overline{ vw} )_{yy} - ( W^2 - V^2 + \overline{w^2} - \overline{v^2} )_{yz} +  \Delta \Delta \Psi/R~.\label{eq:MPSI}
\end{equation}
This vortex is forced by the torque arising from  perturbation Reynolds stresses as discussed in the previous section.  The eddy torque is the only forcing  maintaining the roll circulation against dissipation 
as the mean terms  in (\ref{eq:MPSI}) produce only advection of  streamwise vorticity in the cross-stream/spanwise plane.

Equations (\ref{eq:pv}, \ref{eq:peta}, \ref{eq:MU}, \ref{eq:MPSI}) comprise the roll/streak/turbulence
dynamics.   In the absence of FST this system has as a stable equilibrium solution only the perturbation stable Couette flow 
$U=y$, $V=W=0$.   In the presence of FST  the combined roll/streak/turbulence dynamics includes the ensemble mean Reynolds stress from the perturbation field described by  (\ref{eq:pv}, \ref{eq:peta}, \ref{eq:MU}, \ref{eq:MPSI}) giving rise to  new equilibria.  We next show how SSST can be used to find these new stable equilibria. 

\section{The SSST system governing roll/streak/perturbation dynamics}

The SSST system\citep{Farrell-Ioannou-2003-structural} includes the three components of the streamwise mean   flow  
(\ref{eq:MU}, \ref{eq:MPSI}), and the  ensemble mean Reynolds stress from the  perturbations (\ref{eq:pv}, \ref{eq:peta}).  In the perturbation equations (\ref{eq:pv}, \ref{eq:peta}) stochastic excitation is introduced to parameterize both the exogenous FST and the endogenous scattering by eddy-eddy interactions.   The perturbation equations  with this  parameterization comprise the Stochastic Turbulence Model (STM). 
The STM provides accurate eddy structure at the energetic scales 
because the highly non-normal dynamics associated 
with the non-normal linear operator in strongly sheared flows predominates  in determining the  perturbation structure
\citep{Farrell-Ioannou-1998b, Laval-etal-2003}.
This parameterization  has been widely used  to  describe
the dynamics of turbulence in channel flows
\citep{Farrell-Ioannou-1993e, Farrell-Ioannou-1994b, Farrell-Ioannou-1998a, Bamieh-Dahleh-2001, Jovanovic-Bamieh-2005, Hoepffner-Brandt-2008,
Hwang-Cossu-2010, McKeon-etal-2010} and 
has also been instrumental in  advancing robust control of channel flow  turbulence \citep{Farrell-Ioannou-1998b,  Bewley-Liu-98,  Kim-Bewley-2007,  Hogberg-etal-03a, Hogberg-etal-03b}.    
The STM has also been verified to
determine with great accuracy the eddy structure of the midlatitude  atmosphere \citep{Farrell-Ioannou-1995, DelSole-Farrell-1996,
DelSole-96,  Whitaker-Sardeshmukh-98, Zhang-Held-99, DelSole-01a, DelSole-04}.

We use Fourier expansion in the streamwise direction, $x$, for the  perturbations that deviate from the streamwise mean: 
\begin{equation}
v= \sum_k \hat v_{k}(y,z,t) e^{i k x}~,~\eta= \sum_k \hat \eta_{k}(y,z,t) e^{i k x}~,
\end{equation}
in which the $k=0$  streamwise wavenumber is excluded. 
We discretize the perturbation equations (\ref{eq:pv}, \ref{eq:peta})  in the cross-stream, $y$, and spanwise, $z$,  directions with the state $\hat  \phi_k= [ \hat v_k , \hat \eta_k]^T$
prescribed by the  normal velocity and vorticity on a  $y-z$ grid for each $x$ Fourier component.  Streamwise mean Reynolds stresses can be obtained from the streamwise mean covariance matrix of the perturbation state.   Under the ergodic assumption this streamwise mean covariance is the same as the ensemble mean covariance\footnote{Discussion of application of the ergodic assumption in this context with examples of convergence can be found in FI2003\citep{Farrell-Ioannou-2003-structural}.}, 
$\C_k= < \hat \phi_k  \hat  \phi_k^{\dagger} >$ ( in this expression $<\cdot >$ denotes ensemble averaging and the subscript indicates that the statistics are those of the eddy field components with streamwise wavenumber $k$).

In order to evolve the perturbation covariance
we must first specify an excitation to maintain the FST.  We take
 an excitation in (\ref{eq:pv}, \ref{eq:peta})   that is  delta correlated in time and of the general  form:

\[
\left(
\begin{array}{c}
    F_v\\
    F_{\eta}
\end{array}
\right)
=\F \xi
\]
where $\xi(t)$ is a  normally distributed  independent random column vectors
of length equal to twice the number of discretization points, that satisfies:
\[
< \xi_i (t) \xi_{ j} (s) > = \delta_{ij}  \delta(t-s),
\]
with the structure matrix $\F$  determining
the spatial coherence of the forcing of the cross-stream velocity and  vorticity
i.e. the fluid is excited delta correlated in time with a spatial  structure that is a superposition 
of the columns of  $\F$.
In the ensemble equations for the quadratic eddy covariance at streamwise wavenumber $k$ the stochastic forcing enters through its  covariance $\Q_k$:
\begin{equation}
\Q_k=\F \F^{\dagger}~.
\label{eq:Q}
\end{equation}

The qualitative features of the SSST dynamics are insensitive to the structure of the 
forcing, $\Q_k$, as long as it excites 
the most energetic structures.  The reason is the high non-normality of the perturbation operator which 
leads to strong amplification of a few optimal structures that in turn determine the perturbation field\citep{Farrell-Ioannou-1993e, Farrell-Ioannou-1994b}.

We take $\Q_k$ to be proportional to $\M_k^{-1}$, where  $\M_k$ is the metric that determines the perturbation kinetic energy at streamwise wavenumber $k$  through the inner product:
\begin{equation}
E_k = \hat \phi_k^{\dagger} \M_k \hat  \phi_k~.
\end{equation}
This forcing covariance excites the system so that each degree of freedom receives equal energy.  The energy metric is
given by\renewcommand{\L}{\mathbf{L}}

\begin{equation}
\M_k = \frac{1}{4} \left ( \L_u^{k \dagger} \L_u^k + \L_v^{k \dagger}  \L_v^k + \L_w^{k \dagger} \L_w^k  \right )~,
\end{equation} 
where $\hat u_k= \L_u^k \hat \phi_k$,  $\hat v_k = \L_v^k \hat \phi_k$, and   $\hat w_k= \L_w^k\hat  \phi_k$.
Explicitly, the linear operator $\L_v^k$ is the projection,
\begin{equation}
\L_v^k = [ \I ~~\bf 0]~,
\end{equation}
while the two other linear operators are obtained using Equation (\ref{eq:wudiag}):
\begin{eqnarray}
&\L_u^k  =  
\left(
\begin{array}{cc}
 -i k \Delta_h^{-1} \partial_y  &  0   \\
 0  &        -\Delta_h^{-1} \partial_z
 \end{array}
\right)~~,
~~
\L_w^k  =  
\left(
\begin{array}{cc}
 -\Delta_h^{-1} \partial_{yz}^2  &  0   \\
 0  &        -i k \Delta_h^{-1} 
 \end{array}
\right)~.
\end{eqnarray}

The ensemble averaged covariance evolves according to the deterministic Lyapunov equation     \citep{Farrell-Ioannou-1996a}:

 \begin{equation}
  \frac{d \C_k }{d t}~ =~\A_k( U ) \C_k~+\C_k \A_k^{\dagger}( U )~+~f^2 \Q_k~,
   \label{eq:Lyap1}
  \end{equation}  
  in which $f^2$ is an amplitude factor  and   $\A_k(U)$ 
  is the linear  operator  in
  (\ref{eq:pv}, \ref{eq:peta}) which depends on the  streamwise flow $U(y,z,t)$.  In matrix form the operator $\A_k$ in  (\ref{eq:Lyap1}) is: 
\begin{eqnarray}
\A_k(U) & = &
\left(
\begin{array}{cc}
 \L_{OS} &  \L_{C_1}   \\
 \L_{C_2}  &        \L_{SQ} 
\end{array}
\right)~,
\end{eqnarray}
with 
\begin{subequations}
\label{eq:whole}
\begin{eqnarray}
\L_{OS} =  \Delta^{-1} \left (   -i k U \Delta + i k (U_{yy}-U_{zz}) - 2 i k U_z \partial_z - 2 i k (U_z \partial_{yyz}^3 + U_{yz} \partial_{yz}^2)\Delta_h^{-1}+ {\Delta^2}/{R}   \right ) ,
\label{subeq:1}\\
\L_{C_1}  =  2 k^2 \Delta^{-1} \left (  U_z \partial_y + U_{yz}  \right ) \Delta_h^{-1} ,
\label{subeq:2}\\
 \L_{C_2} =  U_z \partial_y - U_y \partial_z  -U_{yz} + U_{zz} \partial_{yz}^2 \Delta_h^{-1} ,
\label{subeq:3}\\
 \L_{SQ}   =    -i k U \Delta + i k U_{zz} \Delta_h^{-1} + {\Delta}/{R} 
\label{subeq:4}
\end{eqnarray}
\end{subequations}

The covariances, $\C_k$,  evolved by  (\ref{eq:Lyap1}) provides the Reynolds stresses for the mean flow equations  (\ref{eq:MU}, \ref{eq:MPSI}).
For example the  Reynolds stress $\overline{u v}$  is given by: 
\begin{equation}
\overline{uv} = \frac{1}{2} {\rm Re} \left ({\rm diag}\left ( \sum_{i=1}^n \L_u^{k_i} \C_{k_i} \L_v^{k_i \dagger}\right )\right )
\end{equation}
where
${\rm diag}$ denotes the  matrix diagonal and $n$ the number of streamwise harrmonics.  
All the Reynolds stresses can be written similarly as linear functions of the covariance matrix and
the streamwise mean equations (\ref{eq:MU}, \ref{eq:MPSI}) can then be expressed concisely in the form:
\begin{equation}
\label{eq:MEAN}
\frac{ d \Gamma}{d t} = G( \Gamma) +  \L \C
\end{equation}
where $\Gamma \equiv [U,\Psi]^T$ denotes the streamwise mean  flow,
$G$ a function of the mean flow that includes the dissipation and external forcing,
$\C=[\C_{k_1},\cdots,\C_{k_n}]$ and 
$\L \C$ is  the forcing of the mean  by the Reynolds stresses,  with $\L$ a linear operator. 

Equations [(\ref{eq:Lyap1}), (\ref{eq:MEAN})] comprise the SSST system for the roll/streak/turbulence dynamics:
\begin{subequations}
\label{eq:SSST}
\begin{eqnarray}
\frac{d \C_k}{dt}  & =&~\A_k( \P \Gamma ) \C_k~+\C_k \A_k^{\dagger}( \P \Gamma )~+~f^2 \Q_k~,\\
 \frac{ d\Gamma}{d t}  &= & G( \Gamma) + \L \C ~,
 \end{eqnarray}
\end{subequations}
with  $\P$ the projection of $\Gamma$ onto the mean streamwise flow so that $\P \Gamma = U$. The SSST dynamics can be written
concisely as
\begin{equation}
\label{eq:SSST_concise}
\frac{d \chi}{dt} = S(\chi)~,
\end{equation}
 by defining the SSST state  $\chi= [\C, \Gamma]^T$.  The equilibrium states satisfy $S(\chi_{eq})=0$.

Equation (\ref{eq:SSST_concise}) constitutes a closed, deterministic, autonomous, nonlinear system for the co-evolution of 
the streamwise mean flow and its consistent field of turbulent eddies. 
Although the effects of the  turbulent fluxes are
retained in this system, the fluctuations associated with
turbulent  eddy dynamics are suppressed so that the dynamics of turbulent eddy/mean flow interaction and particularly the equilibria arising from this interaction are revealed with clarity.
These nonlinear equilibrium states are intrinsically associated with the turbulence and are therefore dynamically distinct from coherent nonlinear states that have been extensively studied in geophysical  \citep{Branstator-etal-1989, Marshall-Molteni-1993} and in  laboratory shear flows\citep{Nagata-1990, Waleffe-2003, Gibson-etal-2008}.

The SSST system is globally stable \citep{Farrell-Ioannou-2003-structural} and the attractor of the SSST system may be  a fixed point,  a limit cycle,  or a chaotic attractor. 
Examples of each of these behaviors has been found in the SSST description of  geophysical and plasma turbulence\citep{Farrell-Ioannou-2003-structural,Farrell-Ioannou-2008-baroclinic,Farrell-Ioannou-2009-plasmas}.

The concept of the SSST system trajectory is novel because it is not the trajectory of a realization of the turbulent system but rather the trajectory of the statistical mean state of the turbulence which evolves on the time scale of the  mean flow.

The SSST system introduces a new stability concept to fluid dynamics which is the stability of an equilibrium between
a mean flow and its consistent field of turbulence.  This stability theory generalizes the traditional hydrodynamic stability theory of Rayleigh\citep{Rayleigh-1880}.  If a mean flow is perturbation unstable (in the sense of Rayleigh) it is also structurally  unstable (in the sense  of SSST).   However, the converse is not true and perturbation stability does not imply structural stability.  In fact, emergence of roll/streak structures in shear flow will be shown to occur in association with structural instability of a perturbation stable state.

\section{Stability analysis of roll/ streak/ turbulence equilibria}

Assume that  for the  given forcing covariance, $f^2 \Q_k$,  the 
equilibrium $\chi_{eq}= [\C_{eq},\Gamma_{eq}]^T$ of the SSST equations (\ref{eq:SSST}) has been determined.
We can study its stability by  linearizing  the SSST system about  this equilibrium. 
The  perturbation equations  take the form
\begin{subequations}
\label{eq:deltaSSST}
\begin{eqnarray}
\frac{d \delta \C_k}{dt}  & =&~\A_k( \P \Gamma_{eq} ) \delta \C_k~+\delta \C_k \A_k^{\dagger}( \P \Gamma_{eq} )~+~\delta \A_k \C_{k eq}~+\C_{k eq}  \delta \A_k^{\dagger}~~,\\
 \frac{ d \delta \Gamma}{d t}  &= & \left . \frac{\partial G }{\partial \Gamma}\right |_{\Gamma_{eq}} \delta \Gamma +  \L \delta \C~.
 \end{eqnarray}
\end{subequations}
where  $\delta \Gamma= [\delta U, \delta \Psi]^T$ is the perturbation in the mean flow quantities: the streamwise flow, $\delta U$,  and the  roll streamfunction, $\delta \Psi$.   The perturbation to the operator, $\A_k$, that controls the eddy field is $\delta \A_k$.   This operator perturbation is produced  by perturbation to the mean streamwise flow, $\delta U$. 
Setting $ \delta \chi \equiv [\delta \C, \delta \Gamma ]^T$ the perturbation equations can be written concisely as:
\begin{equation}
\label{eq:Ls}
\frac{d \delta \chi}{ d t} = \Ls\delta \chi~
\end{equation}
The linear operator $\Ls \equiv  \partial S / \partial \chi |_{\chi_{eq}}$ depends on the equilibrium state $\chi_{eq} =[\C_{ eq}, \Gamma_{eq}]^T$ 
which  in turn depends on the Reynolds number $R$, the channel geometry, the mean flow forcing, and the stochastic excitation 
$f^2 \Q_k$.
Eigenanalysis of  the linear operator $\Ls$  then determines the  structural stability of this equilibrium roll/streak/turbulence complex\footnote{The stability
operator $\Ls$ is discussed more fully in Farrell\&Ioannou\citep{Farrell-Ioannou-2003-structural}.}.

The familiar laminar Couette flow equilibrium 
$\Gamma_{eq}=[U_{eq}=y$,  $\Psi_{eq}=0]^T$ and $\C_{eq}=0$ is a solution
of the SSST equations (\ref{eq:SSST})  in the absence of  FST ($f=0$).  Because at zero forcing the eddy covariance vanishes,  $\C_{ eq}=0$, the first of the perturbation SSST equations (\ref{eq:deltaSSST}) reduces to an unforced Lyapunov equation which inherits the perturbation stability of Couette flow at all Reynolds numbers\citep{Romanov-73}, i.e. the stability of $\A(\P \Gamma_{eq})$.  The second equation  is asymptotically  unforced and clearly stable so in the absence of FST the system (\ref{eq:deltaSSST}) is structurally stable as well as perturbation stable.  From this argument it is clear that structural instability of a flow  that is perturbation stable in the sense of Rayleigh  requires non-vanishing
$\C_{eq}$ or equivalently non zero values of $f$.

In the presence of spanwise homogeneous FST there is a class of spanwise independent equilibria $\Gamma_{eq}=[U_{eq}(y),\Psi_{eq}=0]^T$  with  non vanishing  $\C_{ eq}$.
In these equilibria the mean streamwise
flow  $U_{eq}(y)$ is maintained by a balance between diffusion and the component of 
Reynolds stress divergence $-(\overline{uv} )_y $ in the inhomogeneous (cross-stream) direction.   The equilibria are 
possible because all
the Reynolds stresses are independent of $z$, and  symmetry requires that $\overline{vw}=0$.
These spanwise independent ensemble equilibria in the presence of FST correspond 
to boundary layer flow equilibria that depart from the Couette profile in $y$ but have no $z$ dependence.  
We will demonstrate that for sufficient amplitude of FST these 
equilibria, while remaining perturbation stable,  become structurally unstable giving rise to roll circulations with associated
streaks.

For convenience the amplitude of the forcing, $\Q_k$, is chosen to maintain RMS perturbation velocity $1\%$  of the mean Couette flow velocity when it is used to excite the Couette flow so that when this excitation is introduced into the perturbation variance equation as $f^2 \Q_k$  the adjustable amplitude, $f$, corresponds approximately to  RMS  FST as a percentage of the mean flow velocity.   RMS perturbation velocity is very nearly linearly proportional to $f$ as $f$ increases prior to the bifurcation to roll /streak equilibria although deviating slightly because the mean flow profile deviates from Couette as FST increases (cf. Fig. \ref{fig:6}).  

 We  demonstrate the structural instability of the spanwise independent equilibria $\Gamma_{eq}=[U_{eq}(y),\Psi_{eq}=0]^T$ in the presence of  FST concentrated in a single wavenumber $k$. 
 This simple case reveals the character of the instability and the results do not change qualitatively when multiple wavenumbers of FST are included. We first examine an example at $R=400$, wavenumber $k=1$,  and spanwise periodic channels on the interval  $3/20< L_z / (2\pi) <3/10$.
These channels are narrower than the minimal channel of  Hamilton, Kim and Waleffe\citep{Hamilton-etal-1995} (herafter HKW) for which $L_z/(2\pi)=6/10$.
The calculations were performed with $N_y=21$ and $N_z=20$ points. Convergence was verified by repeating the calculations  at higher resolutions. 
 
\begin{figure*}
 
\includegraphics[width=7in]{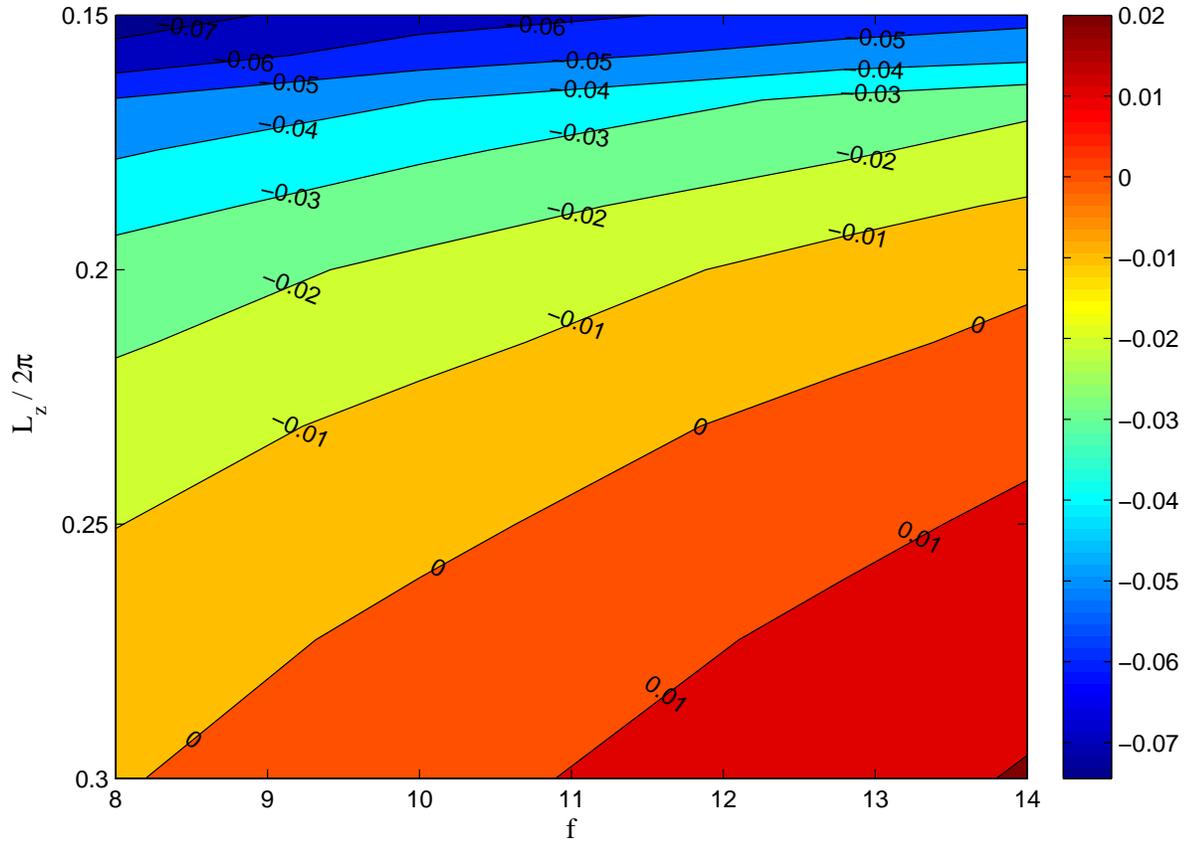}
 \vspace*{-1mm} 
 \caption{(Color online) 
 Growth rate of the most unstable eigenfunction of the SSST system linearized about the  spanwise uniform 
 equilibrium  as a function of STM excitation parameter, $f$, and spanwise channel width, $L_z$. 
 Channel width  $L_z/(2 \pi) = 3/10$, as used in the example of Fig. \ref{fig:3}, lie on the abscissa of this plot.  
Channels  with spanwise  width  $L_z/ (2 \pi)  < 0.205 $ are stable.  
The perturbation field comprises a single wavenumber, $k=1$. The Reynolds number is $R=400$.
 }
\label{fig:2}
\end{figure*}

We use the power method to find the structure and growth rate of the most unstable eigenmode of the
$\Ls$ operator in (\ref{eq:Ls})   
for the spanwise independent equilibrium flow 
$\Gamma_{eq}=[U_{eq}(y)$, $\Psi_{eq}=0]^T$.
Contours of the growth rate of the most unstable eigenmode 
as a function of  STM excitation parameter   $f$  and channel width
$L_z$ are shown in Fig. \ref{fig:2}.   The maximum growth rate of the $\Ls$ operator
increases with $f$ and for a critical intensity, $f_c(L_z)$, the spanwise
independent flow becomes structurally unstable with eigenmodes in the form of exponentially growing roll/streak structures.  The growth rate decreases with  channel width at constant $f$ and for  sufficiently narrow
channels $L_z / (2 \pi) < 0.205$ the spanwise independent mean flow is structurally stable for all $f$   and no roll/streak equilibrium is supported, in qualitative agreement with the findings of the minimal channel 
simulations  of Jimenez and Moin\citep{Jimenez-Moin-1991}.

\begin{figure*}
\includegraphics[width=7in]{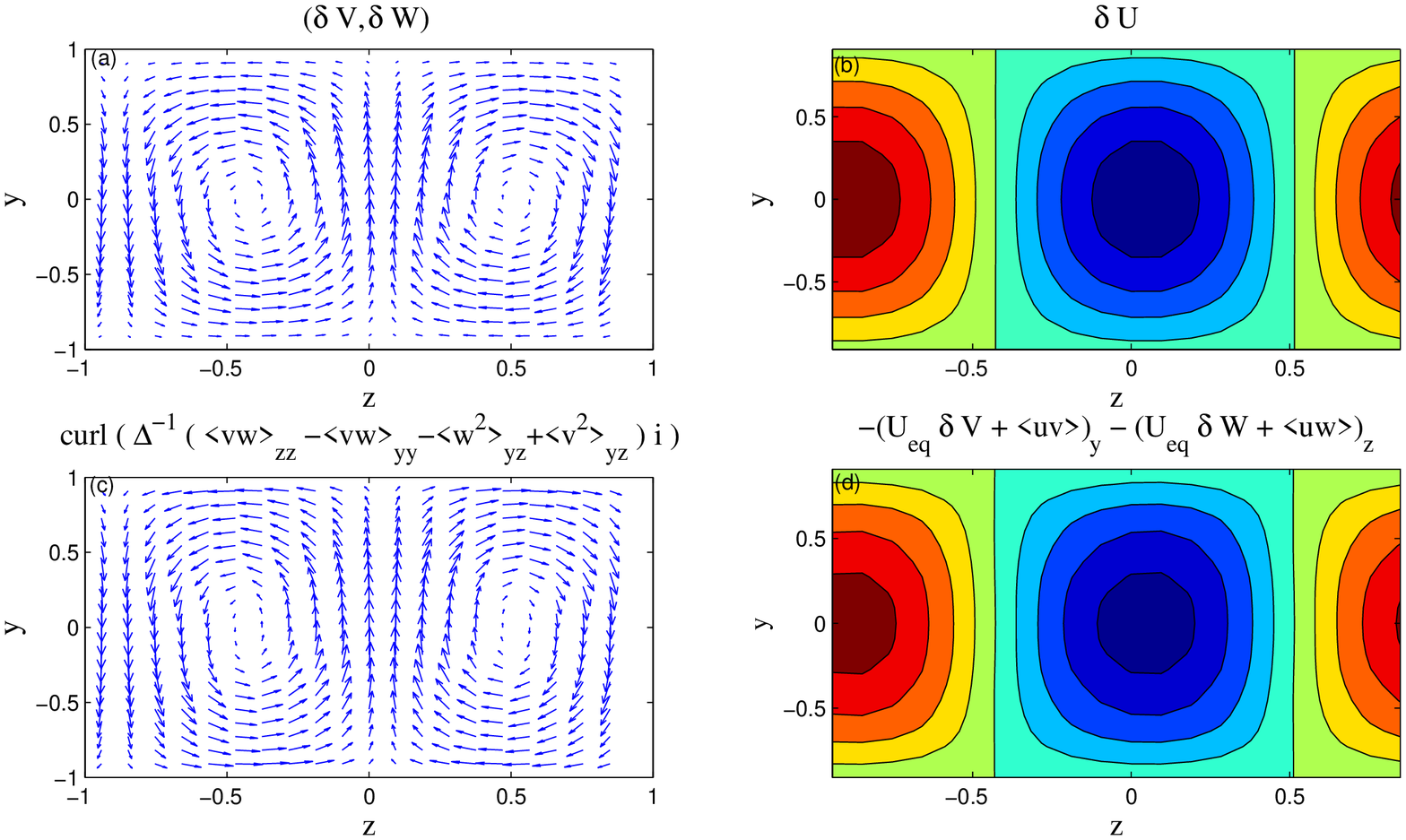}
 \vspace*{-1mm} 
  \caption{(Color online)  The most unstable eigenfunction of the SSST system linearized about the structurally unstable equilibrium with no spanwise variation at STM excitation parameter $f=12.8$.   The growth rate of this eigenfunction is $\lambda= 0.0166$.
 (a):  streamwise mean  cross-stream/spanwise velocity vectors $(\delta V, \delta W)$ in the 
cross-stream/spanwise plane.  (b):  streamwise mean streamwise velocity $\delta U$ associated with the same eigenfunction (negative values dashed).
The ratio of  the maxima of the fields $(\delta U, \delta V, \delta W)$ is $(1,0.06,0.03)$.
The unstable eigenfunction also has  a perturbation covariance component, $\delta C$, the effect of which is indicated by  the  acceleration these perturbations induce in the corresponding velocities (cf.  equations (\ref{eq:MU}) and (\ref{eq:MPSI})): 
(c): $(\delta \dot V , \delta \dot W)$. (d):  $\delta\dot U$.   Parameters are $k=1$, $Lz/(2 \pi )= 0.3$ and $R=400$.}
\label{fig:3}
\end{figure*}

\begin{figure*}
 
\includegraphics[width=7in]{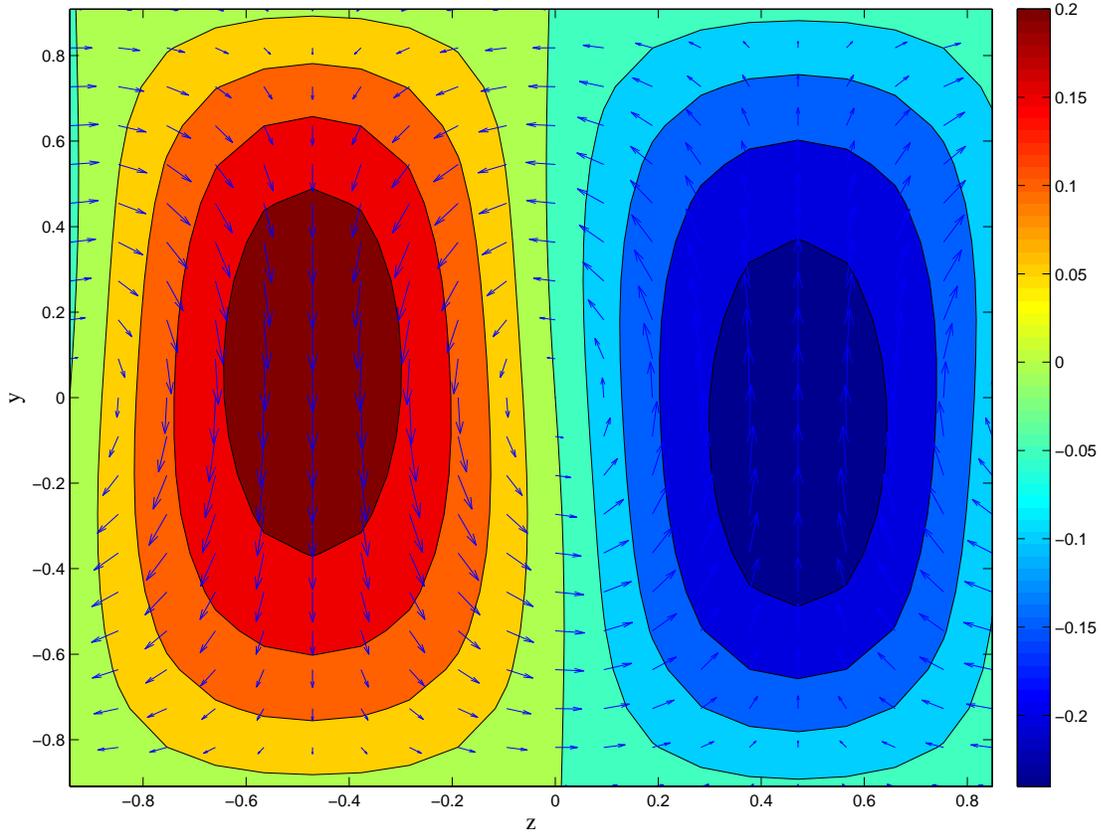}
\vspace*{-1mm} 
\caption{ 
(Color online) Finite amplitude equilibrium  roll/streak structure at STM excitation parameter $f=12.8$.  Shown are contours of 
the streak velocity,  $U_s$, and velocity vectors $(V, W)$ of the roll circulation that
equilibrates from the most unstable eigenfunctions shown in Fig. \ref{fig:3}. }
\label{fig:4}
\end{figure*}

\begin{figure*}
 
\includegraphics[width=7in]{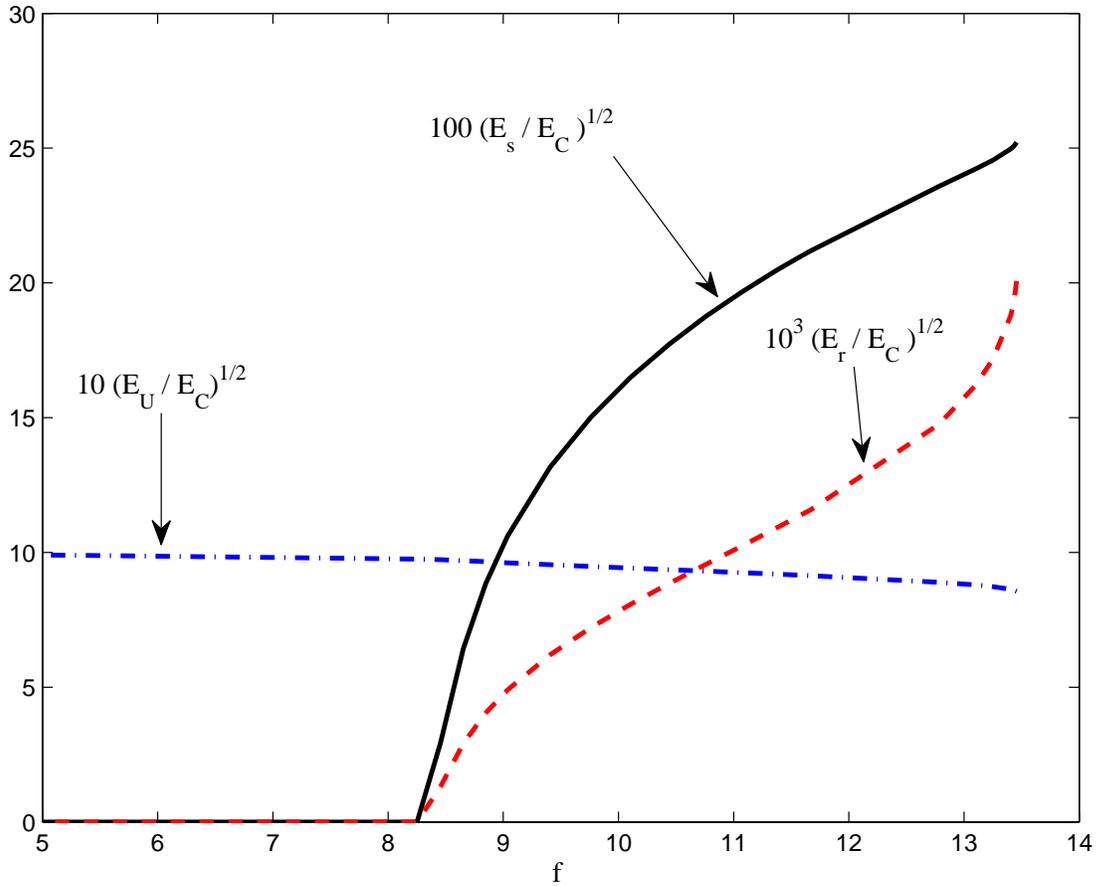}
\vspace*{-1mm} 
\caption{ 
(Color online) Roll/streak equilibria  as a function of  the  STM excitation parameter, $f$.  At the critical  turbulence level corresponding to $f_c=8.25$  the spanwise independent   equilibria bifurcate to spanwise dependent roll/streak equilibria.
Shown are RMS streak  amplitude  (continuous),
RMS roll amplitude  (dashed),  and RMS streamwise  mean flow amplitude (dash-dot)
normalized by the RMS velocity of the unperturbed Couette flow, $\sqrt{E_C}$.
The Reynolds number is $R=400$  and the perturbation field wavenumber is $k=1$.
} 
\label{fig:5}
\end{figure*}

\begin{figure*}
 
\includegraphics[width=7in]{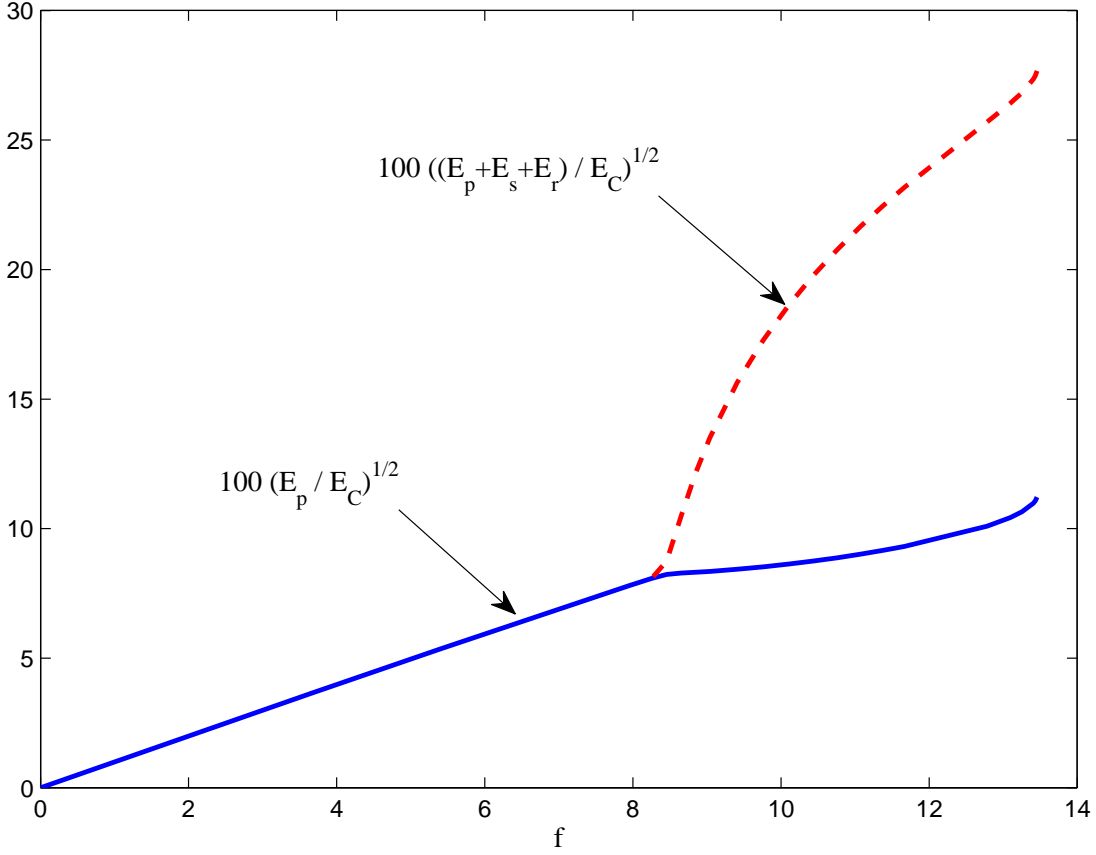}
 \vspace*{-1mm} \caption{(Color online)  
RMS velocity of the  sum of perturbation, roll, and streak components $\sqrt{(E_p+E_m+E_s)/E_C}$ (dashed),
and of the perturbation alone $\sqrt{E_p/E_C}$ (solid) normalized by the Couette RMS velocity, $\sqrt{E_C}$, as a function of STM excitation parameter, $f$.
These curves diverge at the bifurcation STM excitation parameter, $f_c = 8.25$, at which the roll and the streak emerge.  
As the STM excitation parameter increases beyond $f_c$ the roll/streak complex rapidly increases in contribution to the total energy while the perturbation RMS velocity remains near  $10 \%$ of the mean velocity.  Parameters as in Fig.  \ref{fig:3}.}
\label{fig:6}
\end{figure*}

For a channel with $L_z/(2 \pi) = 0.3$ the spanwise  independent equilibrium is structurally 
unstable for  $f>f_c=8.25$.  The growth rate of the most unstable eigenfunction for this channel with  STM excitation parameter
$f=12.8$ is  $\lambda=0.0166$.  The structure  of the
most unstable eigenfunction is shown in Fig. \ref{fig:3}.
 The eigenfunction comprises both a mean flow  perturbation, $\delta \Gamma$, and
an eddy covariance perturbation, $\delta \C$.  The Reynolds stresses associated with $\delta \C$ produce  accelerations and torques in exact 
agreement with the mean flow perturbation consistent with exponential growth.   These structural  instabilities typically equilibrate to finite amplitude roll/streak equilibria similar in structure to the most unstable eigenfunction,  as seen in 
Fig. \ref{fig:4}.
 The critical $f_c$ at which
structural instability occurs  is a bifurcation point in a diagram  of 
equilibria  as a function of  $f$,
as shown in Fig. \ref{fig:5}, \ref{fig:6}.

We denote by $E_r$ the mean  kinetic energy of the roll, obtained by averaging 
$(V^2+W^2)/2$ over the channel; by $E_U$ the mean kinetic energy of the streamwise mean flow, obtained by averaging
$U^2/2$; by  $E_s$ the mean kinetic energy of the streak, obtained by averaging $U_s^2/2$;  and by
$E_p$ the mean kinetic energy of the eddies, obtained by averaging $(\overline{u^2}+\overline{v^2}+
\overline{w^2})/2$. 
As $f$ increases the STM excitation parameter rises and the increasing Reynolds stress 
induces departure of the cross-stream flow from the pure Couette flow but  for $f<8.25$ the mean flow
remains  uniform in the spanwise direction   
and these spanwise independent equilibria 
are stable fixed points of the SSST system.   
The RMS velocity of the perturbation field $\sqrt{E_p/2}$ is shown as a function of $f$  in Fig. \ref{fig:6}.  Note that as the spanwise uniform equilibria bifurcate to roll/streak equilibria for $f>f_c=8.25$ the rate of increase of  perturbation RMS velocity decreases as the turbulence is diverted to drive the roll/streak structure which, being lightly damped, reaches high equilibrium velocity.  It is interesting that the RMS velocity  remains approximately  $10 \%$ of the background flow velocity until forcing excitation $f_u=13.5$ at which point the  roll/streak equilibrium undergoes a second structural instability.  
The roll/streak equilibria  for $f_c < f < f_u$ are perturbation stable, and the breakdown that occurs
at $f_u$ is due to a secondary structural instability of the finite amplitude roll/streak.  This secondary structural instability will be examined in future work.

\begin{figure*}
 
\includegraphics[width=7in]{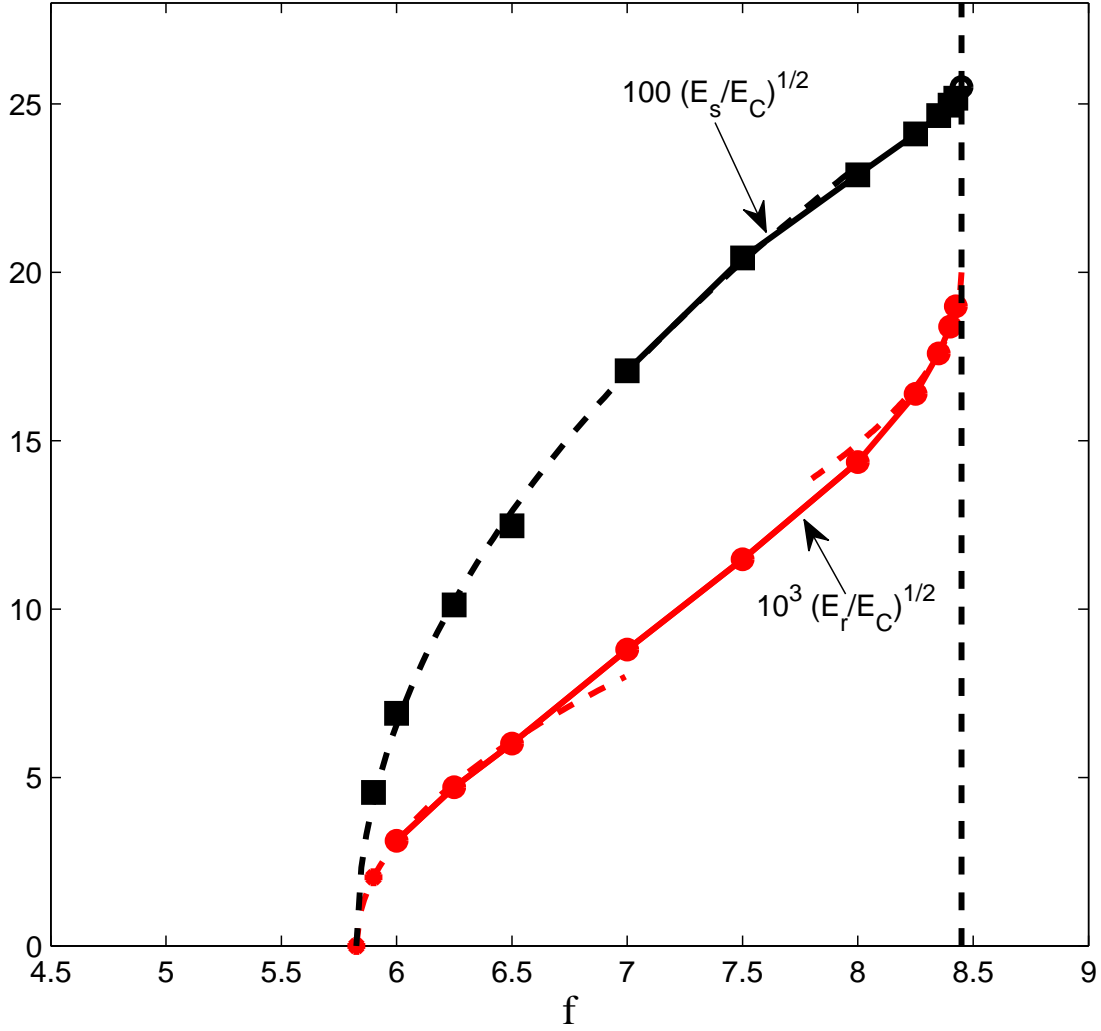}
 \vspace*{-1mm} \caption{ 
(Color online)  Bifurcation diagram of the roll/streak structure as a function of STM excitation parameter, $f$.  At $f_c=5.825$  the spanwise uniform  equilibrium bifurcates to a spanwise dependent roll/streak equilibrium.
Shown as a function of $f$  is normalized streak strength, $100 (E_s/E_C)^{1/2}$ (squares) and normalized roll strength,  
$10^3 (E_r/E_C)^{1/2}$ (circles).   The 
dashed line indicates the $\sqrt{f-f_c}$ dependence of a second order bifurcation.
The stable equilibria extend up to  $f_u=8.45$; at which point the perturbation stable roll/streak equilibrium becomes structurally unstable.  The dashed line indicates the  $\sqrt{f_u-f}$ dependence of a  second order bifurcation.  
Parameters correspond to the  HKW channel:  length $L_x = 1.75 \pi$,  spanwise width $L_z =1.2 \pi$,  half cross-stream height
$L_y = 1.0$ and the Reynolds number is $R=400$.  The perturbation streamwise wavenumber, $k=1.143$, corresponds to the gravest mode in the channel.}
\label{fig:7}
\end{figure*}

\begin{figure*}
 
\includegraphics[width=7in]{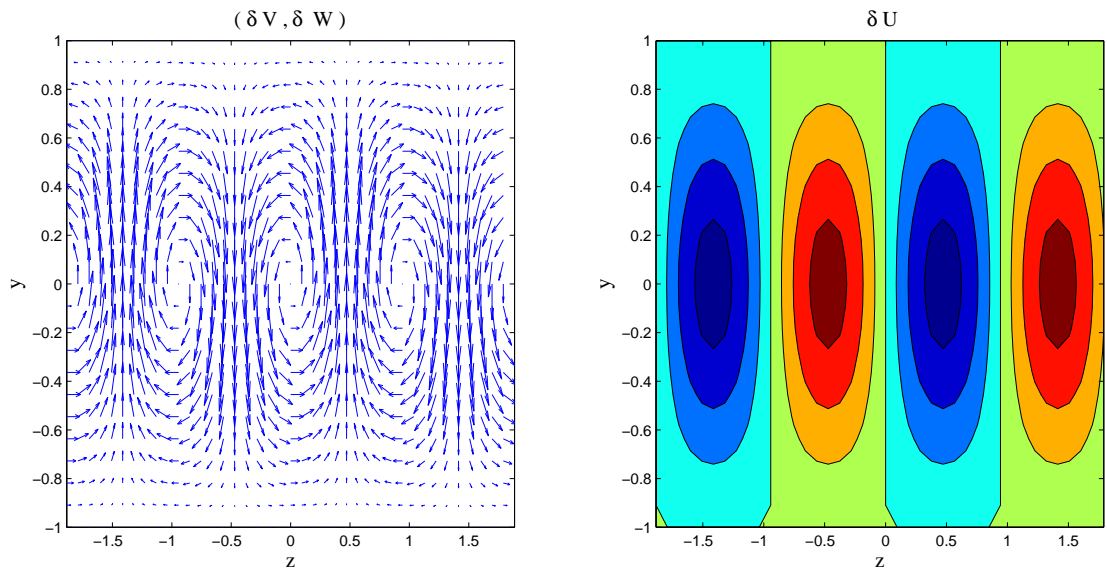}
 \vspace*{-1mm} 
  \caption{(Color online)  The most unstable eigenfunction of the SSST system linearized about the structurally unstable equilibrium with no spanwise variation at STM excitation parameter  $f=8.4$.   The growth rate of this mode is $\lambda= 0.014$.
Left:  streamwise mean  cross-stream/spanwise velocity vectors $(\delta V, \delta W)$ in the 
cross-stream/spanwise plane. Right:  streamwise mean streamwise velocity $\delta U$.
The ratio of  the maxima of the fields $(\delta U, \delta V, \delta W)$ is $(1,0.06,0.03)$.
Parameters correspond to the  HKW channel:  length $L_x = 1.75 \pi$,  spanwise width $L_z =1.2 \pi$,  half cross-stream height
$L_y = 1.0$ and the Reynolds number is $R=400$.  The perturbation streamwise wavenumber, $k=1.143$, corresponds to the gravest mode in the channel.
}
\label{fig:8}
\end{figure*}

We next demonstrate structural instability of the Couette flow in the minimal channel  considered by HKW\citep{Hamilton-etal-1995} taking $L_z/(2 \pi) =0.6$, $R=400$ and  the  smallest streamwise wavenumber in the HKW channel,  $k=1.143$.    These calculations were made with $N_y=21$ and $N_z=40$ points.   This flow bifurcates
from spanwise independent equilibria to  spanwise dependent equilibria 
at  $f_c=5.82$ as shown 
in Fig. \ref{fig:7}.   Note the qualitative similarity with the bifurcation diagram for the
smaller channel shown above.  Both the  RMS streak velocity  and the RMS roll velocity 
vary as $\sqrt{f-f_c}$ near the  the bifurcation point (dashed curve in Fig. \ref{fig:7}) consistent with a supercritical pitchfork.  
At $f_u=8.45$  the roll-streak equilibrium loses structural stability 
and no nearby equilibrium or  periodic solution exists for $f>f_u$ .  For values close to this second 
structural instability  the equilibria exhibit a
$\sqrt{f_u-f}$ behavior consistent with a second order subcritical  bifurcation which will be examined in future work. 
The most unstable perturbation of the SSST system about the unstable equilibrium state without roll/streak structure is shown for STM excitation parameter
$f=8.4 < f_u$  in Fig. \ref{fig:8}.

\section{Structure of the  roll/streak  equilibria}

\begin{figure*}
 
\includegraphics[width=7in]{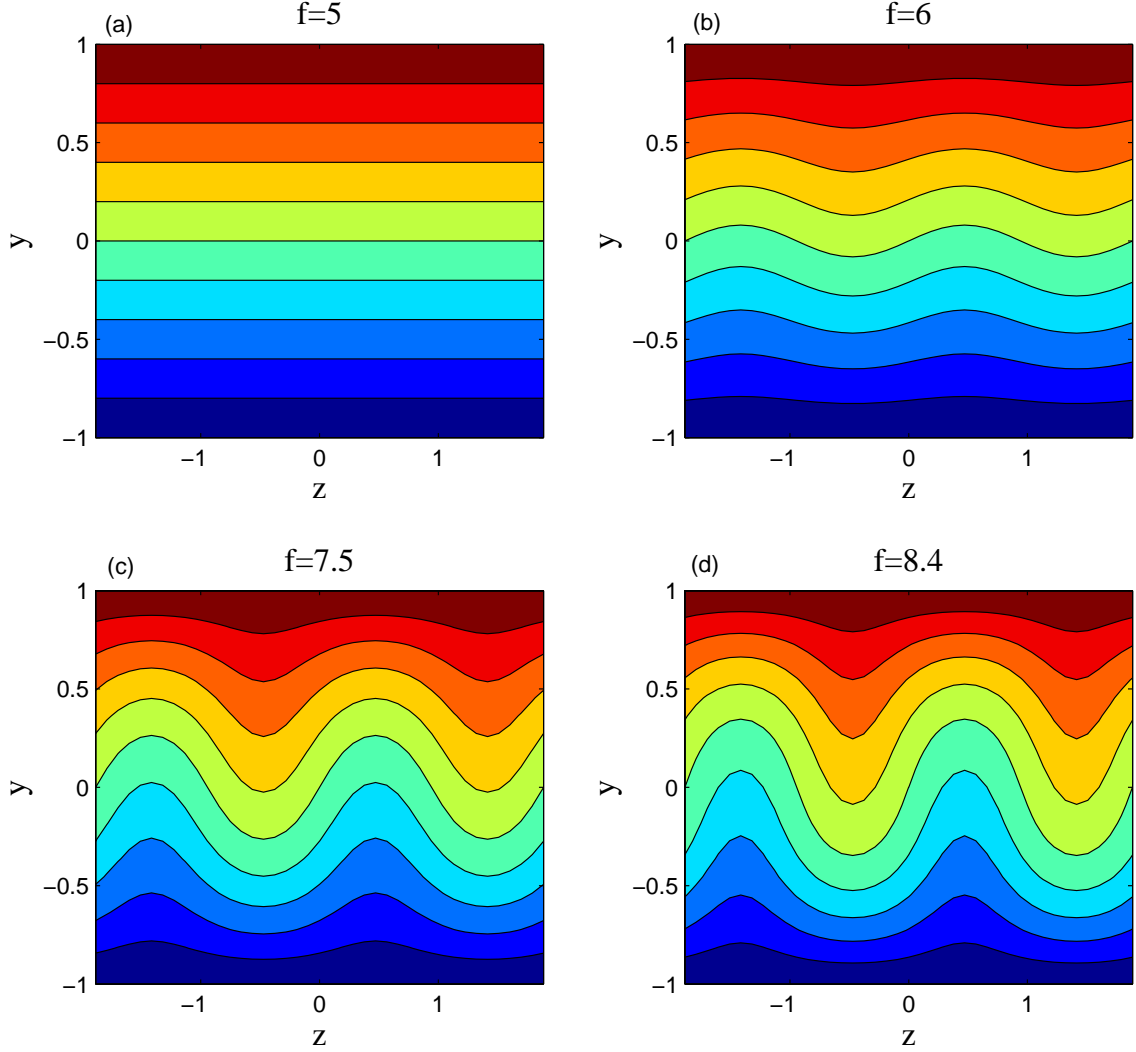}
 \vspace*{-1mm} 
  \caption{(Color online)  Streamwise mean velocity in the $(y,z)$ plane for  the 
 equilibria at various STM excitation parameter values, $f$.   (a): at STM excitation parameter value  $f=5$ the equilibrium is spanwise uniform and there is no associated roll/streak.
 (b): At  $f=6$ the spanwise independent flow is structurally unstable and the associated equilibrium flow has a weak streak with associated roll/streak. (c): The 
 equilibrium at $f=7.5$. (d): The equilibrium at $f=8.4$. All these equilibria are perturbation stable.  Parameters correspond to the  HKW channel:  length $L_x = 1.75 \pi$,  spanwise width $L_z =1.2 \pi$, half  cross-stream height
$L_y = 1.0$ and the Reynolds number is $R=400$.  The perturbation streamwise wavenumber, $k=1.143$, corresponds to the gravest mode in the channel.}
\label{fig:9}
\end{figure*}

\begin{figure*}
 
\includegraphics[width=7in]{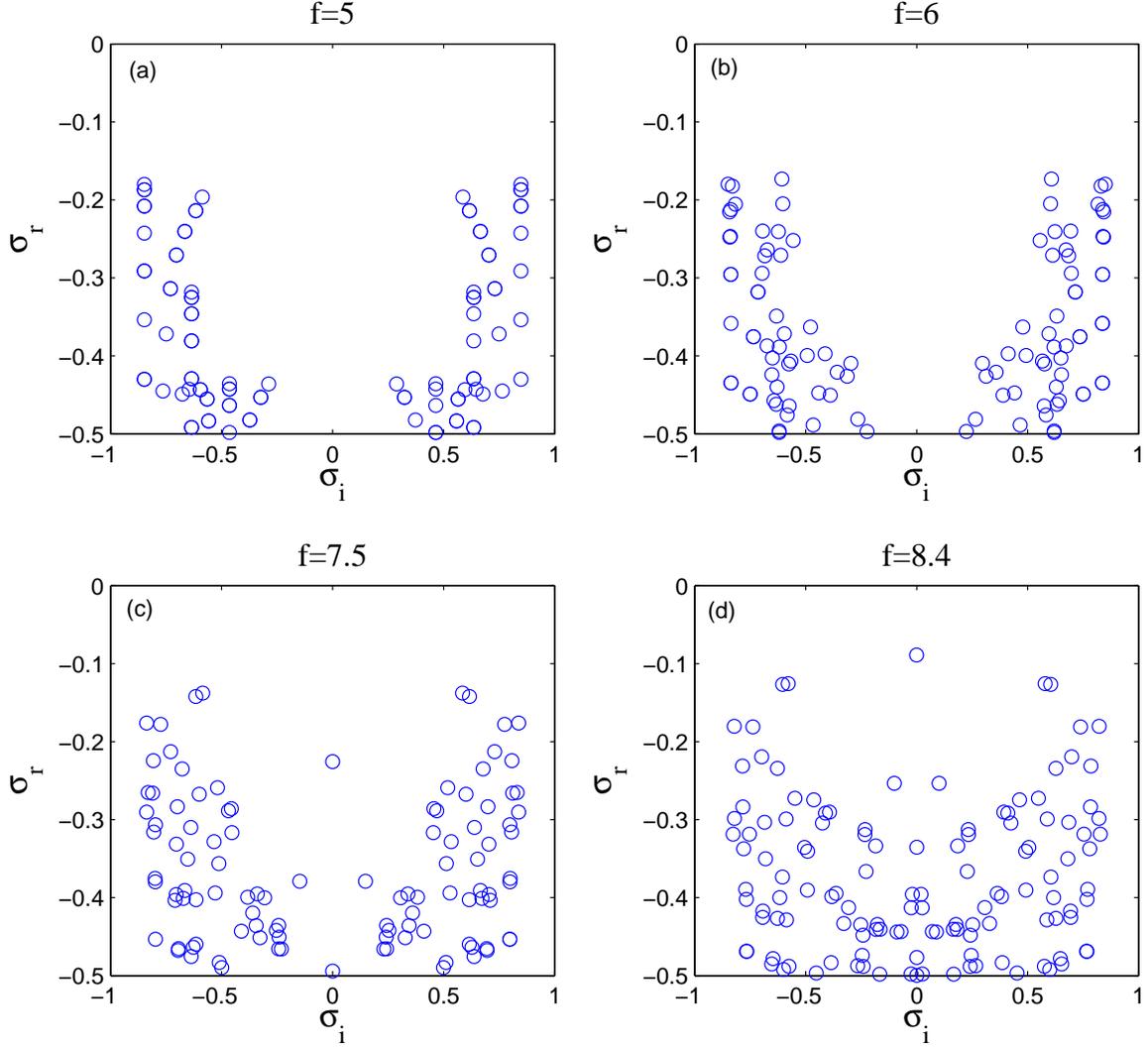}
 \vspace*{-1mm} 
  \caption{(Color online)  The least stable eigenvalues $(\sigma_r,\sigma_i)$ of the operators $\A_k$ that govern the perturbation stability of 
  the equilibrium flows shown in Fig. \ref{fig:9}.  All the flows are perturbation stable. Note the emergence of a mode with $\sigma_r=0$ as the streak increases in magnitude.  Parameters correspond to the  HKW channel:  length $L_x = 1.75 \pi$,  spanwise width $L_z =1.2 \pi$,  half cross-stream height
$L_y = 1.0$ and the Reynolds number is $R=400$.  The perturbation streamwise wavenumber, $k=1.143$, corresponds to the gravest mode in the channel.}
\label{fig:10}
\end{figure*}

The streamwise mean flows of the structurally stable roll/streak 
equilibria in the  HKW  channel  for STM excitation parameter in the interval $f_u  >f>f_c$  are shown in   Fig. \ref{fig:9}.
These equilibria exhibit  streamwise high and low speed streaks that increase in amplitude as $f$ increases. 
A useful measure of streak strength is its lift angle\citep{Schoppa-Hussain-2002}
which we define here as:
\[
\theta_s = \max  \tan^{-1} \left (\frac{\partial_z U}{\partial_y [U]} \right )~,
\]
where $[U]$ is  the spanwise average streamwise mean flow $U$. These equilibrium streaks  reach
$\theta_s = 56^o$ at $f=8.45$. Despite their high lift angles all these equilibria are perturbation stable.
The eigenvalues,  $\sigma$, of the perturbation operators, $\A_k$, for the flows 
in Fig. \ref{fig:9} are shown in Fig. \ref{fig:10}.  Note the emergence  of a mode with frequency $\sigma_i =  0$ as the streak increases in amplitude with increase in $f$. This is the sinuous mode that is associated with the spanwise inflection of the streak and that is often assumed to be responsible for streak breakdown\citep{Hamilton-etal-1995,Waleffe-1995a,Waleffe-1995,  Waleffe-1997, Reddy-etal-1998} .  However 
at $f=8.45$ the roll/streak flow is still robustly stable and the instability  that occurs at  $f=8.45$ is solely a structural  instability of the 
cooperative turbulence/mean flow SSST dynamics.

\begin{figure*}
 
\includegraphics[width=7in]{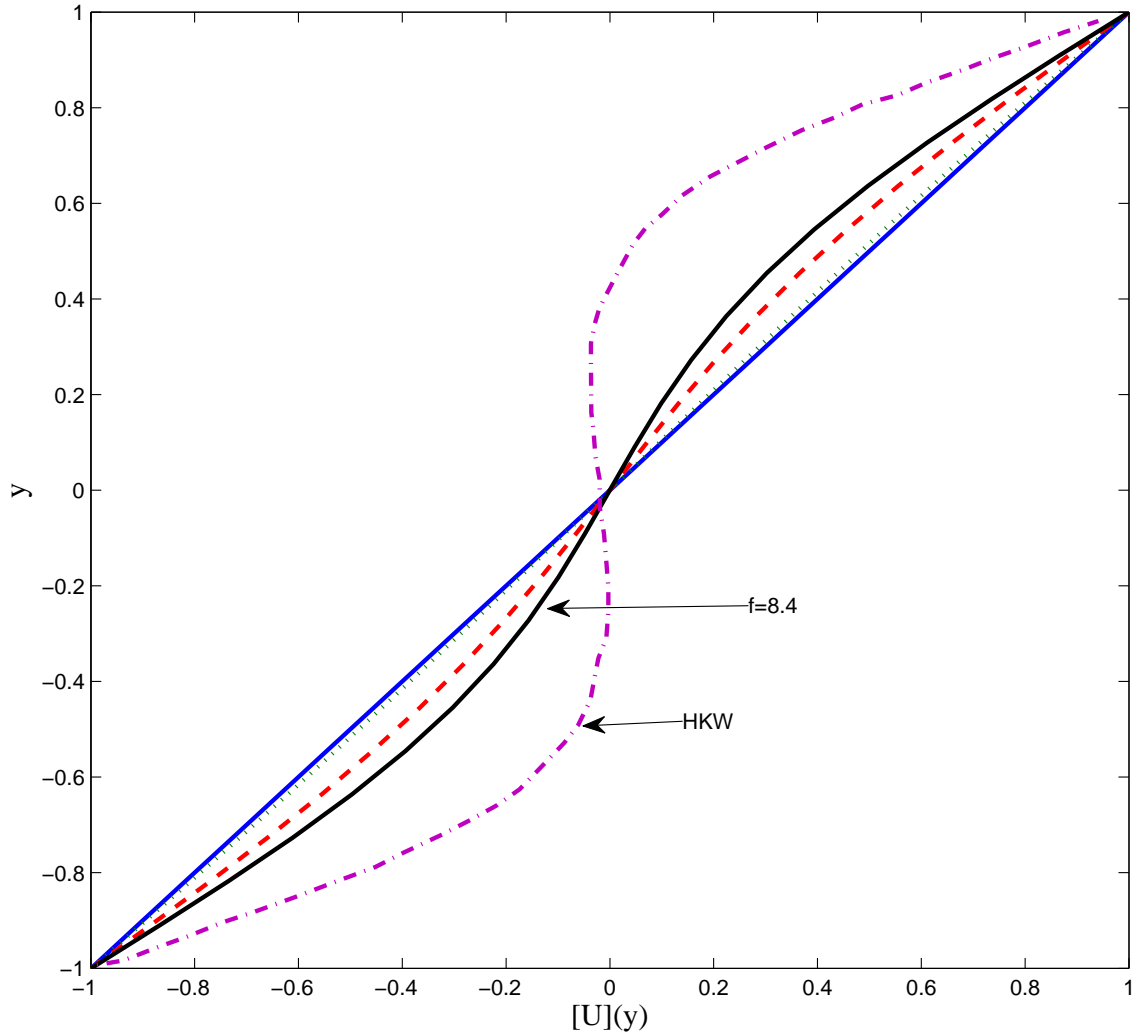}
\vspace*{-1mm} 
\caption{ 
(Color online) The spanwise averaged streamwise flow, $[U](y)$,   for the  SSST equilibria  for STM excitation parameter $f=8.4$ (solid), $f=7.5$ (dashed), $f=6$ (dotted) as shown in Fig. \ref{fig:9}. 
Also shown for comparison is 
the  time and spanwise  mean  under turbulent conditions (dash-dot) as well as the laminar
 Couette flow.  The laminar roll/streak  equilibrium at $f=8.4$   produces dissipation  $1.4 D_C$, where 
 $D_c$ is the dissipation in Couette
 flow, and the half channel  width in wall units at this value of $f$ is $L_y = 24 y^+$.  Parameters correspond to the  HKW channel:  length $L_x = 1.75 \pi$,  spanwise width $L_z =1.2 \pi$,  half cross-stream height
$L_y = 1.0$ and the Reynolds number is $R=400$.  The perturbation streamwise wavenumber, $k=1.143$, corresponds to the gravest mode in the channel.} 
\label{fig:11}
\end{figure*}

Despite the large lift angle the equilibrium streamwise mean flows do not resemble
the mean flows of the turbulent state. In Fig. \ref{fig:11}  we compare
the spanwise averaged streamwise mean flow $[U]$ for the flows with  STM excitation parameter
$f=(8.4, ~7.5,~6)$  with the corresponding time averaged mean flow in the turbulent HKW channel. 
The equilibrium mean flows that occur
in this range of STM excitation parameter  indicate that these roll/streak  equilibria  are laminar.  Indicative of this laminar regime
is the viscous dissipation of the streamwise mean flow: 
\[
D= \frac{1}{R} \int_0^{L_z} dz \int_{-L_y}^{L_y} dy (U_y^2 +U_z^2 +V_y^2 +V_z^2 +W_y^2+W_z^2) ~,
\]
The ratio $D/D_C$, where $D_C$ is the dissipation associated with the Couette flow, of  the equilibria
for FST intensities $8.45 >f>f_c=5.82$ is in the range  $1<D/D_C<1.4$ while this ratio is of order
 $3$ in the turbulent state.  

The laminar roll/streak equilibria  shown in Fig. \ref{fig:9} have  spanwise wavenumber 2.  The spanwise width of the channel is $90 y^+$ for the equilibrium with $f=8.4$ (the wall unit is defined as $y^+ \equiv 1 / \sqrt{ R   [ U_y ] }$, where 
$R$ is the Reynolds number and $[U_y] $ is the mean shear at the boundary) implying streak spacing $45 y^+$ which is half that found in turbulent boundary layers.
However, it should be kept in mind that this wall unit is being calculated for an essentially laminar flow.   In section VIII we show that this spacing does agree with the observed streak spacing of $2$ displacement thicknesses that is observed in laminar boundary layers before transition \citep{Westin-etal-1994}.

\begin{figure*}
 
\includegraphics[width=7in]{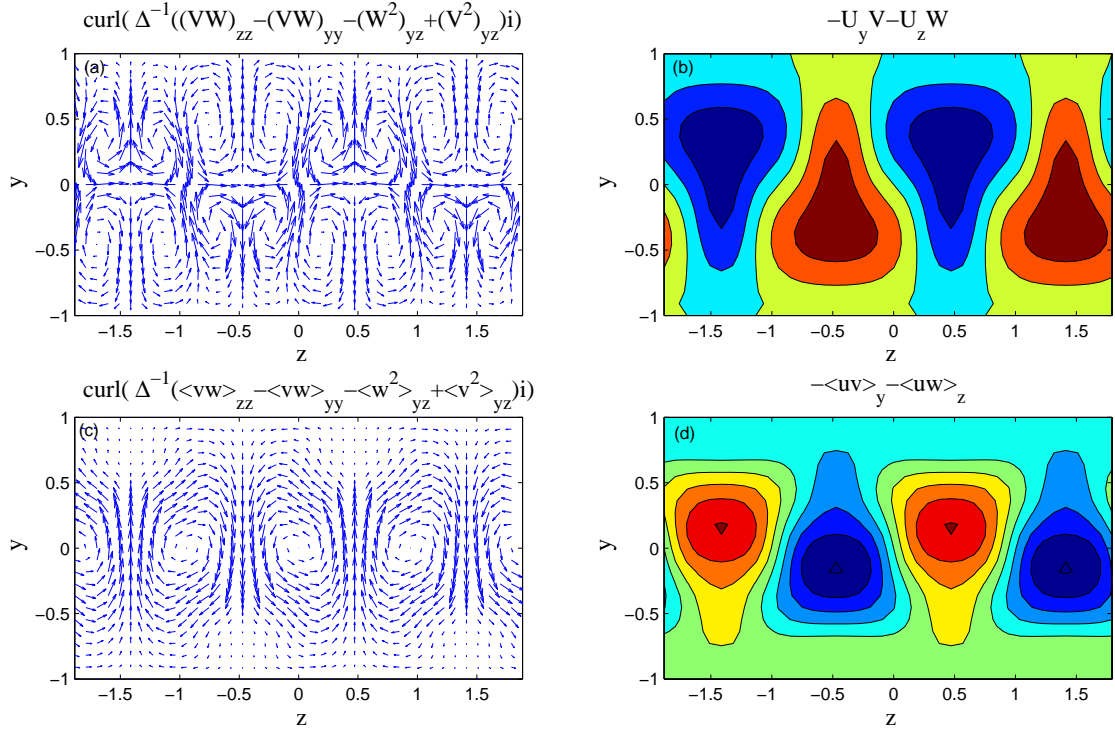}
 \vspace*{-1mm} \caption{(Color online)  
For the equilibrium shown in Fig. \ref{fig:9}d;
 (a): acceleration vectors, $(\dot V, \dot W)$, of the streamwise mean roll circulation 
 induced by the mean velocity 
 field, the maximum  $\dot V$ is $10^{-4}$.
  (b): acceleration of the mean streamwise flow, $\dot U$,   induced by the streamwise mean roll 
 circulation, the maximum  $\dot U$
 is $10^{-2}$, mainly due to the lift-up mechanism.  (c):
 acceleration vectors, $(\dot V, \dot W)$, of the streamwise mean roll circulation 
 induced by the eddy field, the maximum  $\dot V$ is $10^{-3}$. 
 (d): acceleration of the mean streamwise flow, $\dot U$,  induced by the eddy field,  
 the maximum  $\dot U$
 is $10^{-2}$.  The eddy field decelerates the streaks.
Parameters correspond to the  HKW channel:  length $L_x = 1.75 \pi$,  spanwise width 
$L_z =1.2 \pi$,  half cross-stream height
$L_y = 1.0$ and the Reynolds number is $R=400$.  The perturbation streamwise wavenumber, $k=1.143$, corresponds to the gravest mode in the channel.}
\label{fig:12}
\end{figure*}

\begin{figure*}
 
\includegraphics[width=7in]{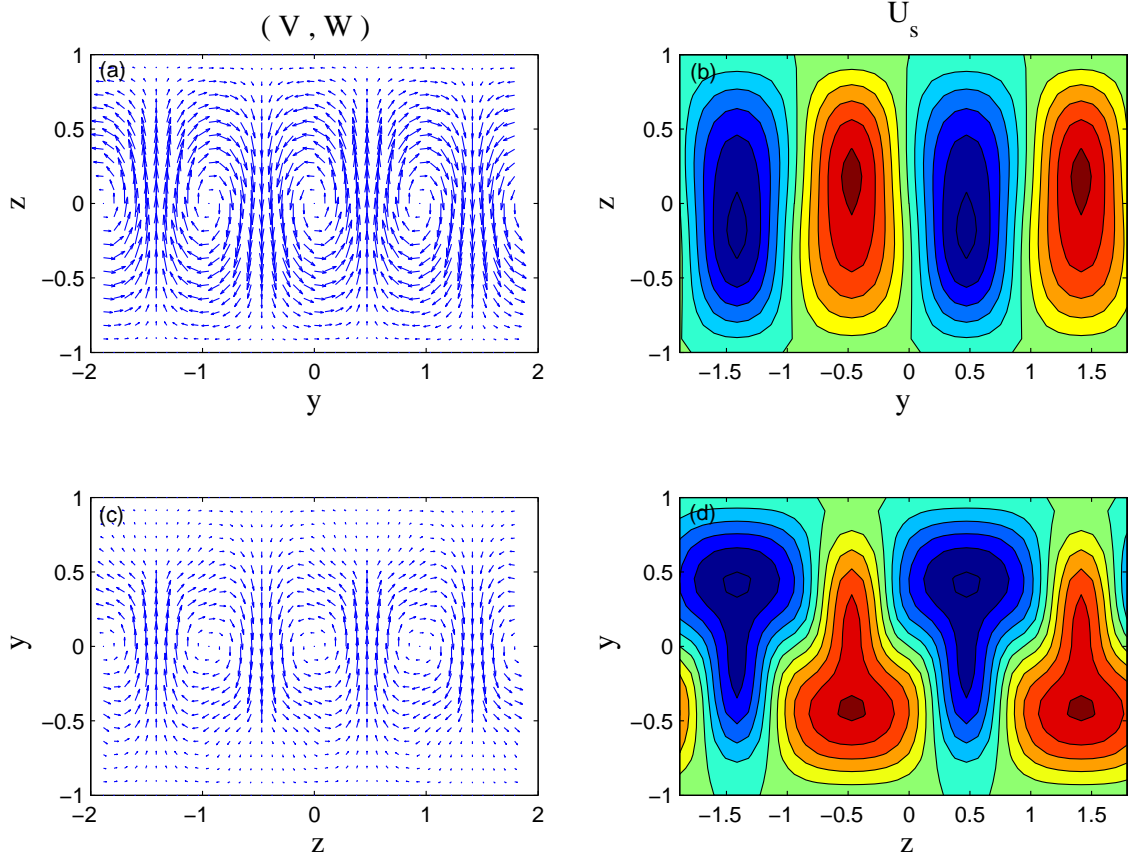}
\vspace*{-1mm} 
\caption{ 
(Color online) 
For the equilibrium shown in Fig. \ref{fig:9}d;
(a): Roll vector velocity $(V, W)$  at equilibrium,  the maximum $V$ velocity is $0.02$ and
the maximum $W$ velocity is $0.009$.  (b):  Streak  velocity $U_s=U-[U]$ at equilibrium,
the maximum velocity is $0.26$.
 (c): mean roll acceleration   vectors     induced by both  the mean
 and eddy fields  given by\\
 $\nabla \times ( \Delta^{-1}((VW+<vw>)_{zz}-(VW+<vw>)_{yy}-(W^2+<w^2>)_{yz}+(V^2+<v^2>)_{yz})i).$  The roll  circulation is maintained against friction by the  eddy field. 
 (d): mean  streamwise acceleration  induced by both  the mean
 and eddy fields $-(UV+<uv>)_y-(UW+<uw>)_z .
 $\\
The mean streamwise flow is maintained against friction by the  lift up mechanism.}
\label{fig:13}
\end{figure*}

Consider the mechanism maintaining the roll/streak structure  at STM excitation parameter $f=8.4$ (Fig. \ref{fig:9}d).
The roll circulation is  maintained against friction only  by the turbulent stress divergence in (\ref{eq:MPSI}) as the quadratic 
streamwise mean terms in (\ref{eq:MPSI}) do not generate mean streamwise vorticity. 
The cross-stream/spanwise acceleration $(\dot V, \dot W)$ due to the eddy flux divergence is shown in Fig. \ref{fig:12}c.  Note that this acceleration is consistent with the circulation shown in Fig. \ref{fig:13}a.  
The acceleration induced by  the  mean momentum  flux divergence is subdominant in these equilibrium solutions
and as a result the total acceleration  $(\dot V, \dot W)$ has the structure of the acceleration induced  by the eddies  as
shown in Fig \ref{fig:13}c and Fig \ref{fig:13}a. 

While the roll circulation is maintained against friction solely by the  torque induced by the
Reynolds stress divergences,  the streak is influenced both by  Reynolds stress divergence and by  mean 
momentum flux divergence 
(cf. Eq. \ref{eq:MU}).  The mean momentum flux divergence can be identified with the 
lift up mechanism as shown in Fig. \ref{fig:12}b 
and this mechanism dominates in the streak maintenance (cf. Fig. \ref{fig:12}b and Fig. \ref{fig:12}d).
The eddy Reynolds stress divergence, shown in Fig. \ref{fig:12}d,  tends to damp the streak  consistent with 
 the eddies  extracting energy from the spanwise shear.  
This mean deceleration  of the streak by the 3-D eddies in laboratory shear flow contrasts with the
acceleration by  quasi 2-D eddies that is primarily responsible for
 jet formation in planetary atmospheres\citep{Jeffreys-1933, Starr-1968, Robinson-1996}.
In compensation for the loss of this dominant  2-D jet formation mechanism, these 3-D  shear flows gain an indirect pathway for maintenance of the streak:  the Reynolds stress
divergences induce roll circulations which through the lift-up mechanism maintain the  streaks  against both viscous dissipation and 
the deceleration induced  by the turbulent Reynolds stress divergence.  
This dual role of the eddy field in maintaining the  equilibrium roll/streak structure  will be discussed further in the 
next section.  

\begin{figure*}
 
\includegraphics[width=7in]{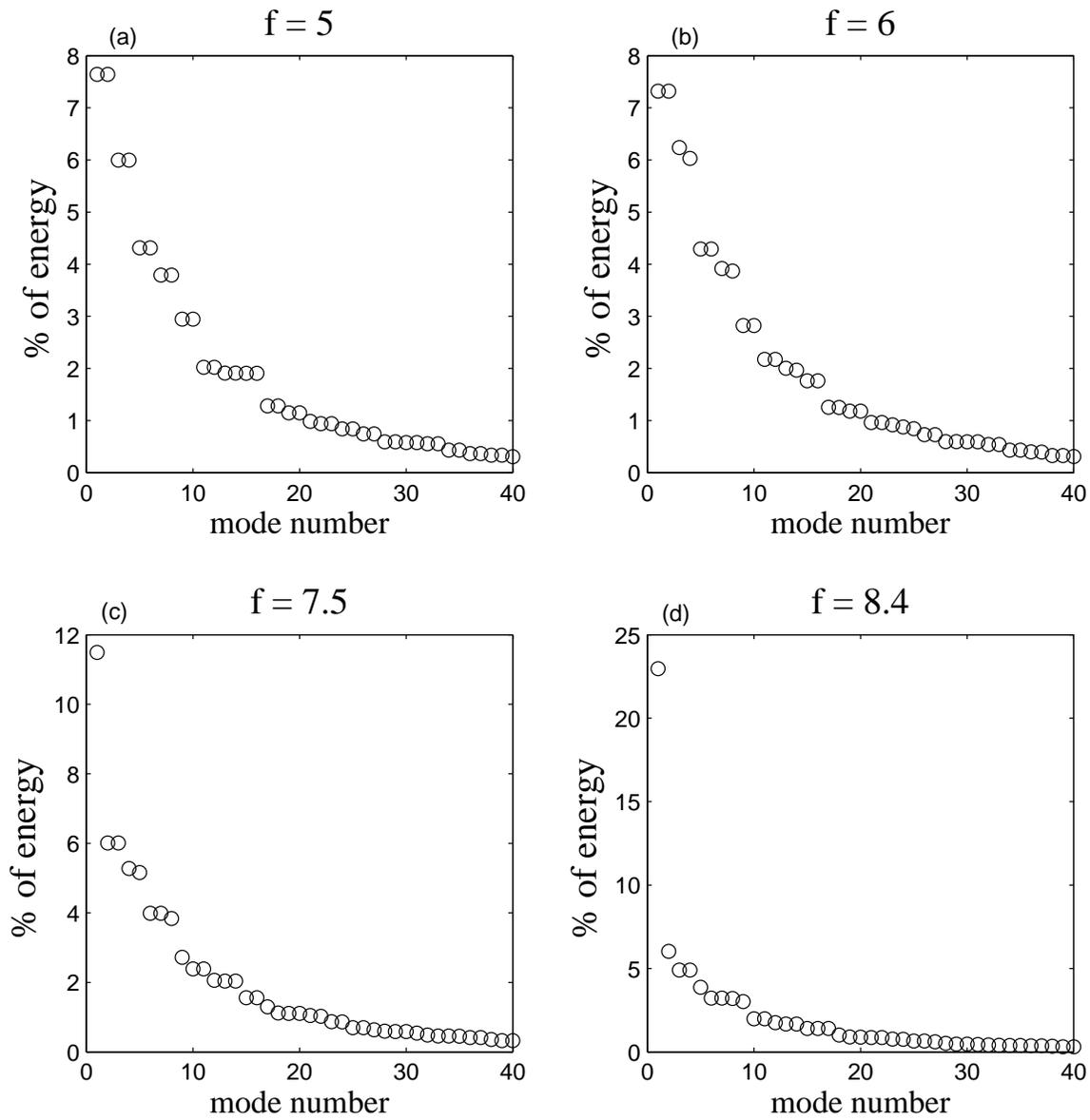}
\vspace*{-1mm} 
\caption{ 
(Color online)  Percentage contribution of the leading EOF's  of the equilibrium covariance to the total  eddy mean energy  maintained by the equilibria shown in Fig. \ref{fig:9}. }
\label{fig:14}
\end{figure*}

\begin{figure*}
 
\includegraphics{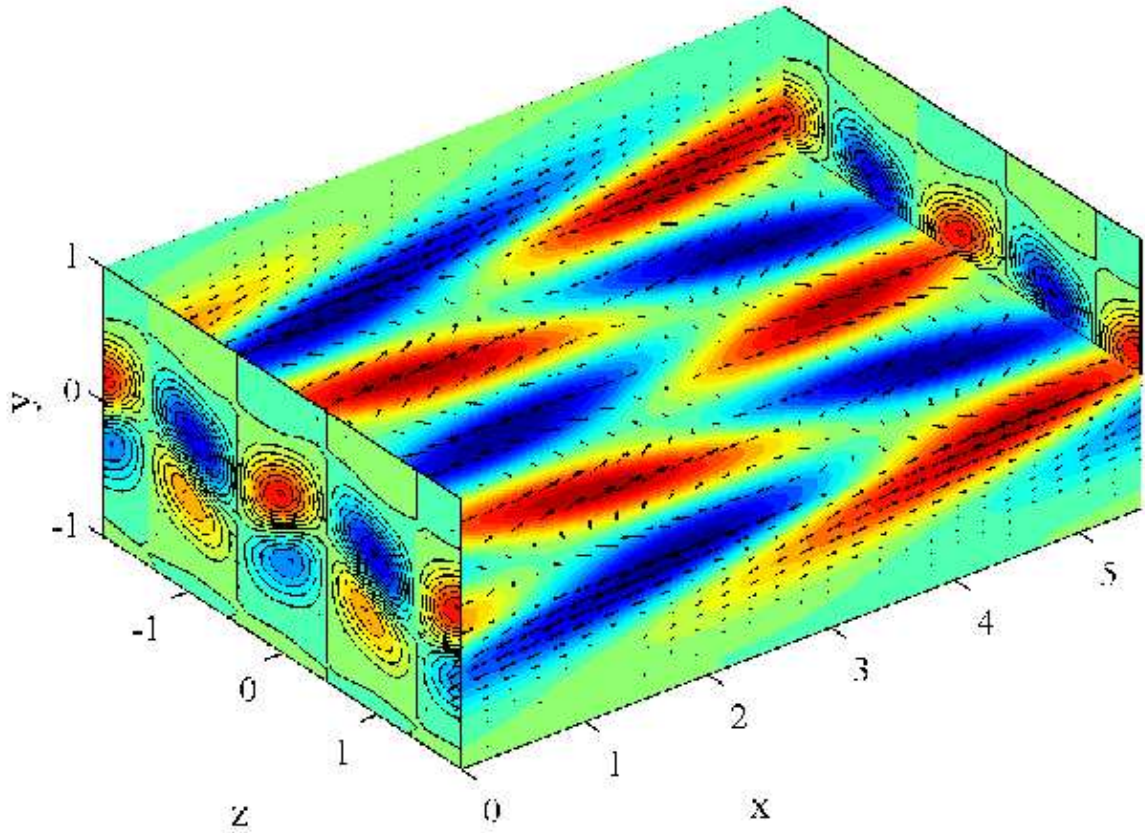}
\vspace*{-1mm} 
\caption{ 
(Color online) Velocity field of the gravest  EOF  
accounting for $24\%$ of the  eddy energy for STM excitation parameter $f=8.4$.
Velocity vectors are shown  with contours of  streamwise velocity.
Parameters correspond to the  HKW channel:  length $L_x = 1.75 \pi$,  spanwise width 
$L_z =1.2 \pi$,  half cross-stream height
$L_y = 1.0$ and the Reynolds number is $R=400$.  The perturbation streamwise wavenumber, $k=1.143$, corresponds to the gravest mode in the channel.
} 
\label{fig:15}
\end{figure*}

\begin{figure*}
 
\includegraphics[width=7in]{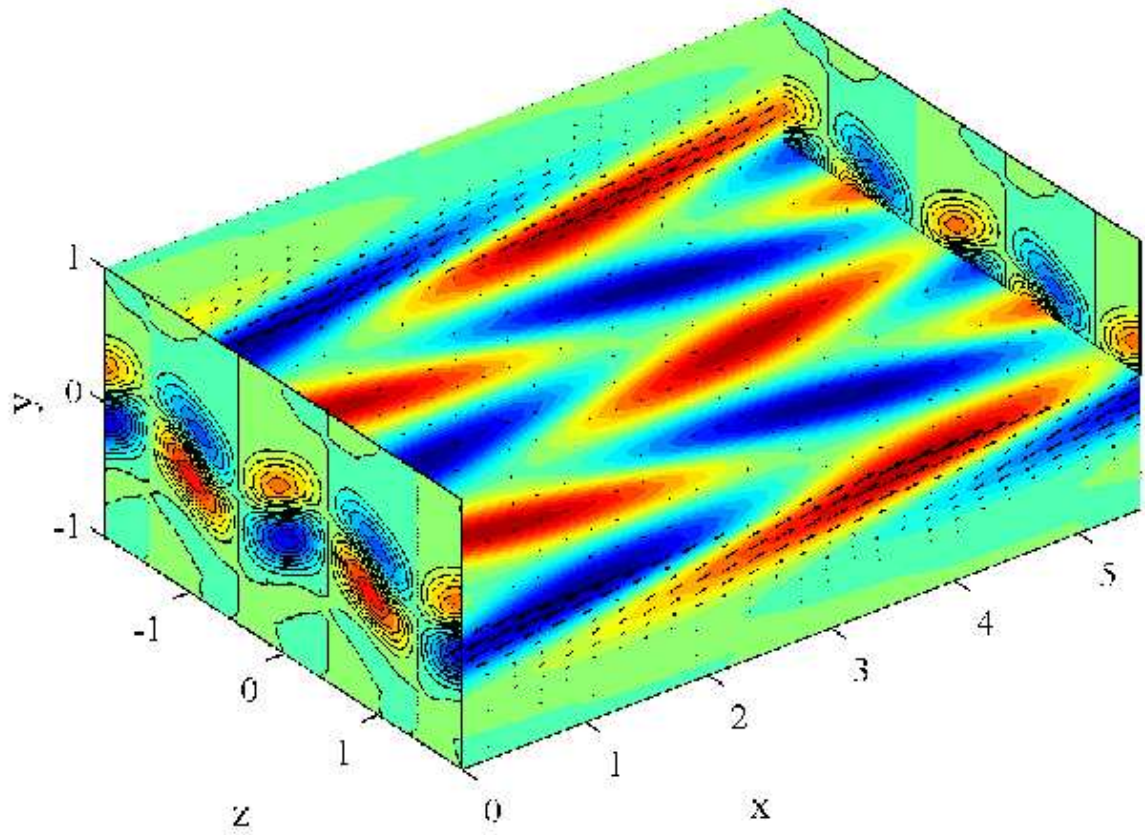}
\vspace*{-1mm} 
\caption{ 
(Color online) 
Velocity field of the least stable mode with eigenvalue ($\sigma_r=-0.017$, $\sigma_i=0$),  for  the equilibrium at STM excitation parameter $f=8.4$.
Velocity vectors are shown  with contours of  streamwise velocity.
Parameters are as in  Fig. \ref{fig:15}.
} 
\label{fig:16}
\end{figure*}

\begin{figure*}
 
\includegraphics[width=7in]{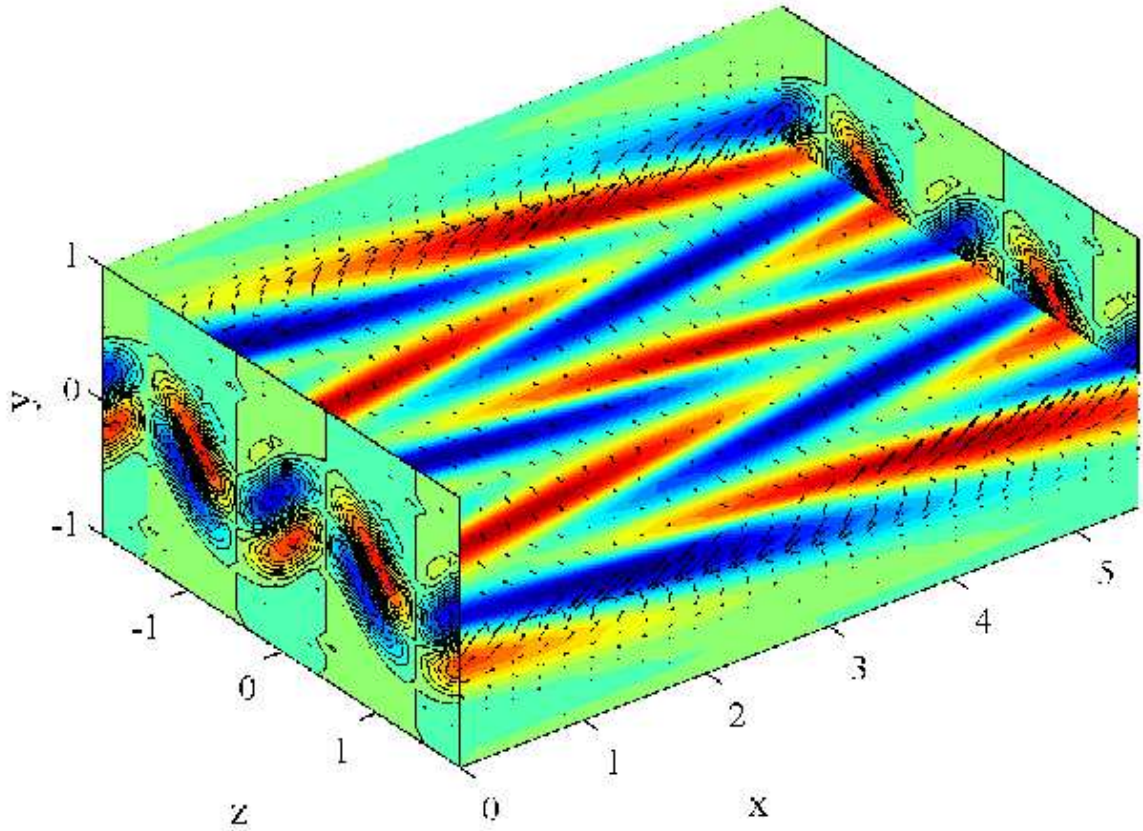}
\vspace*{-1mm} 
\caption{ 
(Color online) Velocity field of the adjoint in the energy inner product of the  least stable mode with eigenvalue ($\sigma_r=-0.017$, $\sigma_i=0$), for  the equilibrium at STM excitation parameter $f=8.4$.  Velocity vectors are shown  with contours of  streamwise velocity. The adjoint is the optimal excitation of the mode. An initial condition consisting of the adjoint with unit energy excites the least stable mode a factor of 1900 greater than an initial condition consisting  of the least stable mode itself with unit energy demonstrating that the mode amplitude derives almost entirely from
non-normal growth processes. 
 Parameters are as in  Fig. \ref{fig:15}.
} 
\label{fig:17}
\end{figure*}

 We turn now to the structure of the eddy field at equilibrium. 
Eddy structures can be ordered in energy by  eigenanalysis of $\M^{1/2} \C \M^{1/2}$, where $\M^{1/2}$ denotes the square root of the  metric $\M$. Eigenfunctions of $\M^{1/2} \C \M^{1/2}$  in descending order of  eigenvalue define
 the empirical orthogonal function (EOF) or Karhunen-Loeve decomposition of the eddy field.  The percentage of the energy accounted for by each of the first  40 gravest modes of the covariance is shown
in Fig. \ref{fig:14} for STM excitation parameter $f=5$, for which the flow is spanwise uniform, and also for  intensity parameters $f=6, ~7.5, ~ 8.4$.  It is clear that the variance is spread over many structures but as $f$
 increases the first EOF begins to dominate and its structure becomes a good representation of the eddy structure. This dominant EOF for the equilibrium structure at  $f=8.4$  is shown  in Fig. \ref{fig:15}. 
 This  eddy field is characterized by  sinuous oblique waves  centered at the wings of the streak and slanted in the vertical.
 This is the  structure that produces the coherent torques maintaining the roll circulation.  The dominant EOF is
 close in structure to  the least stable mode of the flow which is shown 
 in Fig. \ref{fig:16}. Note however that this mode is robustly stable so it is maintained by excitation and non-normality, with the latter dominant.  
 The large excitation of the mode is due to 
 the interaction between this mode and the other modes of the system as revealed by its optimal excitation structure which is its adjoint\citep{Farrell-1988a,Hill-95,Farrell-Ioannou-1996a}. 
 The adjoint of the least stable mode  in the energy inner product is shown in Fig. \ref{fig:17}. 
 An initial condition consisting of its adjoint excites the least stable mode at amplitude
 a factor 1900 greater than an initial condition consisting of the least stable mode itself so the mode arises out of the FST primarily due to excitation of its adjoint.

\begin{figure*}
 
\includegraphics[width=7in]{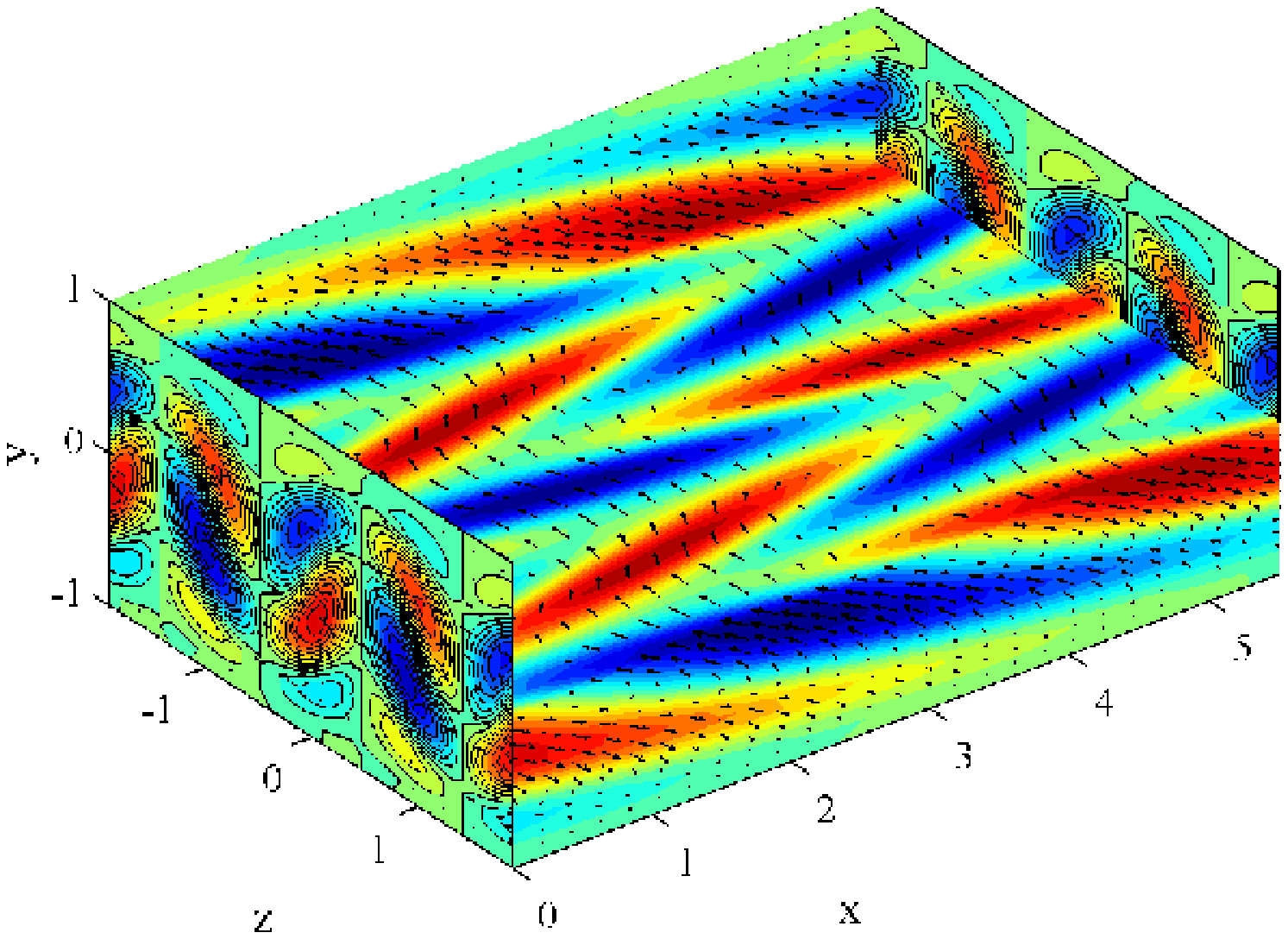}
\vspace*{-1mm} 
\caption{ 
(Color online) Velocity field  at time $t=0$  of the optimal perturbation that maximizes energy growth at $t=10$ 
for the equilibrium flow  with  STM excitation parameter $f=8.4$. 
Velocity vectors are shown  with contours of  streamwise velocity.
Parameters are as in  Fig. \ref{fig:15}.} 
\label{fig:18}
\end{figure*}

\begin{figure*}
 
\includegraphics[width=7in]{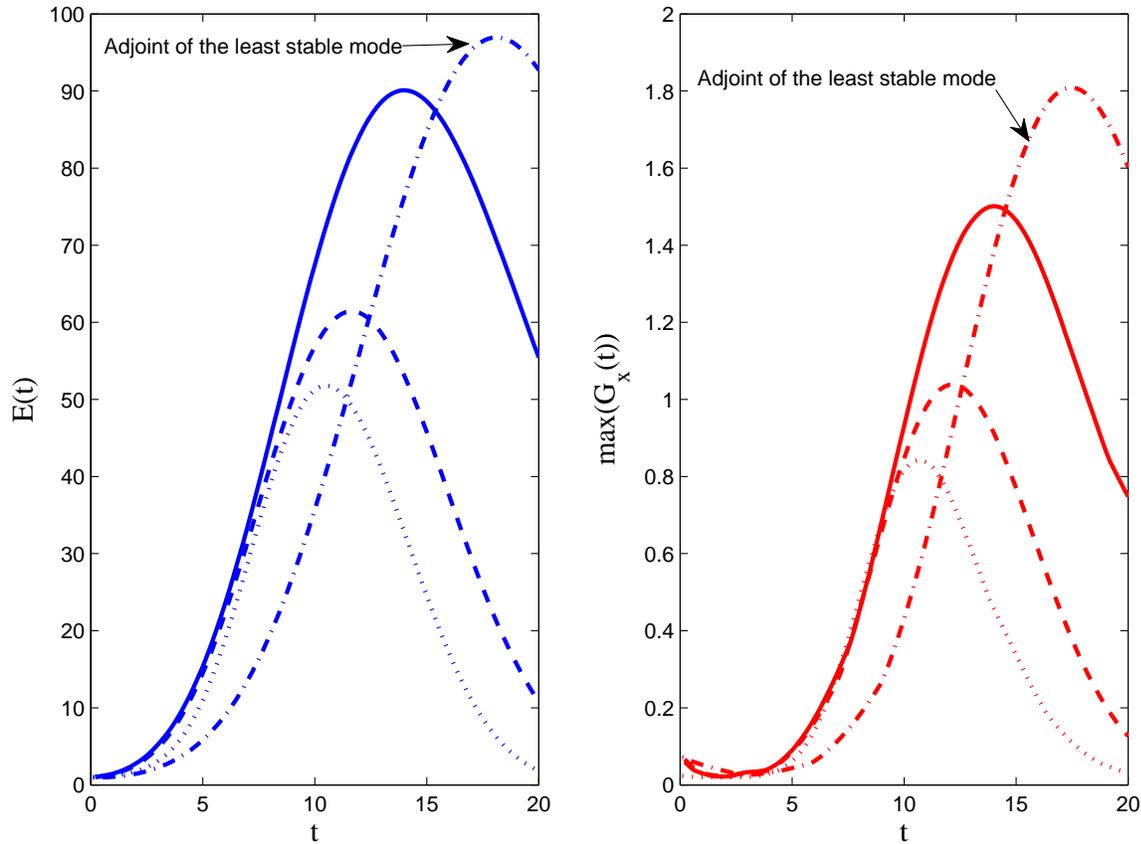}
\vspace*{-1mm} 
\caption{ 
(Color online) Left :  Energy  of the optimal perturbation 
that maximizes energy growth  at $t=10$ as  a function of time for the equilibrium flow  
with  STM excitation parameter  $f=8.4$ (solid),
$f=7.5$ (dashed) and for $f=5$ for which there is no roll/streak; (dotted).
Also shown is the energy growth associated with a unit energy  initial condition
consisting of the adjoint in the energy metric of the least damped  mode for the equilibrium with $f=8.4$ (dash-dot). 
Right: The time development of the maximum  mean streamwise torque induced by the Reynold's stresses 
of the corresponding evolving optimals and the adjoint.  
This figure demonstrates  that both the mode amplitude and its contribution to the streamwise mean 
torque are due to
non-normal growth processes.  All modes in 
these flows are exponentially stable (cf. Fig. \ref{fig:10}a,c,d).
Parameters are as in Fig. \ref{fig:15}.} 
\label{fig:19}
\end{figure*}

\begin{figure*}
 
\includegraphics[width=7in]{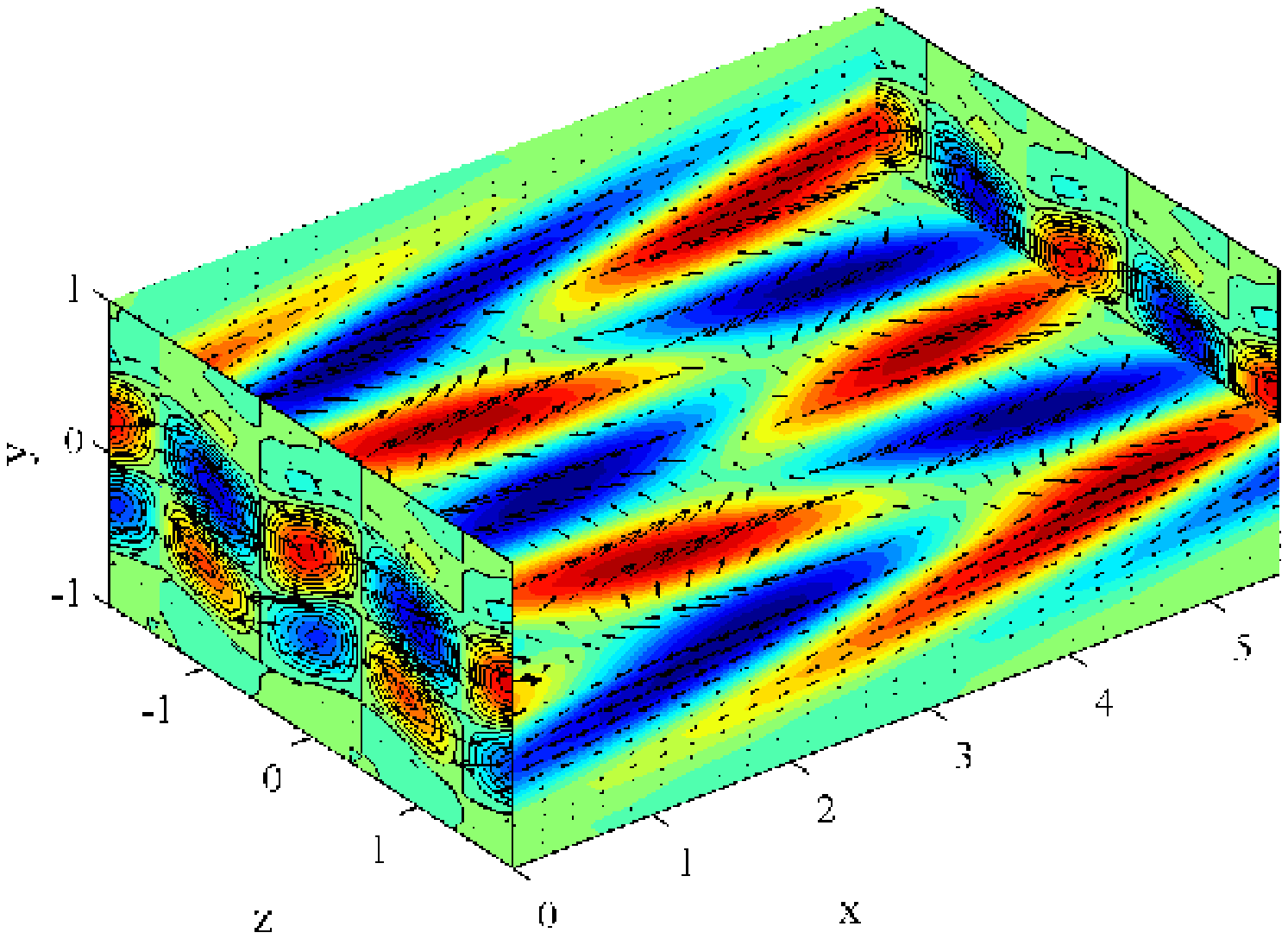}
\vspace*{-1mm} 
\caption{ 
(Color online) 
Velocity field  at time $t=15$  of the optimal perturbation that maximizes energy growth at $t=10$ 
for the equilibrium flow  with  STM excitation parameter $f=8.4$. 
Velocity vectors are shown  with contours of  streamwise velocity. Parameters are as in  Fig. \ref{fig:15}. } 
\label{fig:20}
\end{figure*}

Because the excitation is chosen to be white in energy and all modes are stable the 
structure of the eddy field can be understood dynamically by examining optimal structure evolution.
The optimal perturbation that leads to the greatest growth in energy in 10 time units for the equilibrium 
at  $f=8.4$ is shown in Fig. \ref{fig:18}.  
The energy growth of this optimal is close to the energy growth
of the adjoint of the least stable mode (cf. Fig. \ref{fig:19}).
Evolution of the maximum mean streamwise torque, $G_x$, induced by the
Reynolds stress divergence of these optimal perturbations  is also shown in Fig. \ref{fig:19}.  The torque 
increases as the perturbation energy increases. The energy evolution of the $t=10$ optimal for the equilibria
with $f=7.5$ and $f=5$, for which value  the flow is spanwise independent, are also shown.
 The structure of the optimal when it reaches its maximum energy at $t=15$ is shown in 
 Fig. \ref{fig:20}.  The structure  of this evolved optimal perturbation
 is similar to the structure of evolved optimals  in equilibria at lower FST intensities as 
 expected from the universality of the dynamics of oblique perturbations
 in three dimensional shear
flows\citep{Kelvin-1887b, Moffatt-67, Farrell-Ioannou-1993a,Farrell-Ioannou-1993b}.   The
 spanwise streak serves to collocate the perturbation structures  aligning them
so that the spanwise Reynolds stress divergence produce torques in phase with the evolving roll.

%
 
\begin{figure*}
 
\includegraphics[width=6in]{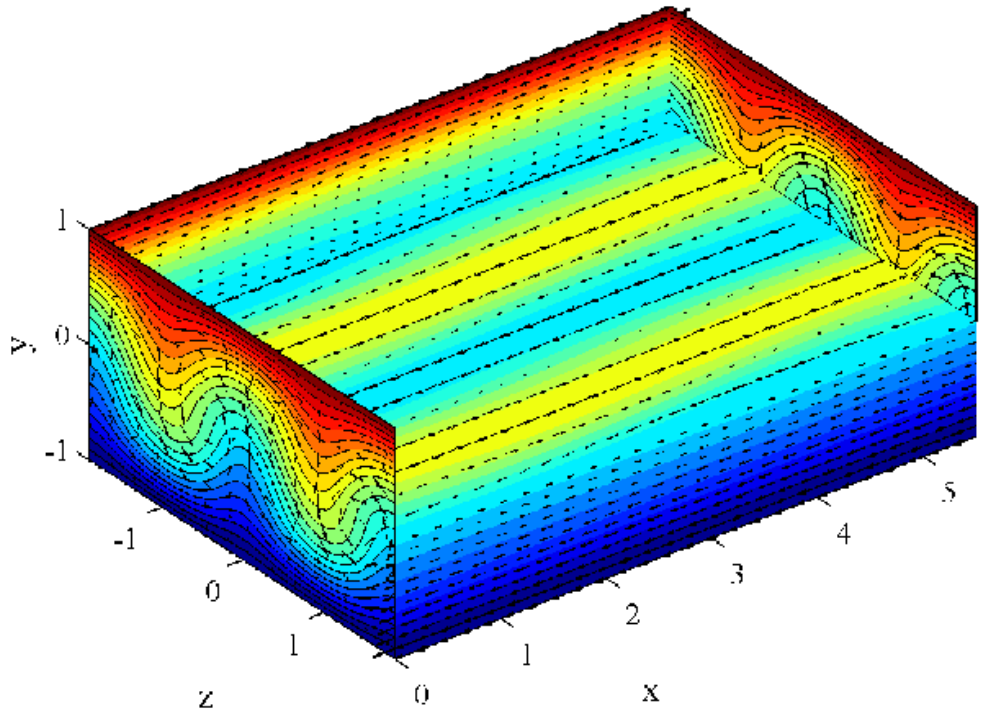}
\vspace*{-1mm} 
\caption{ 
(Color online) Velocity field (mean and perturbation) 
for the equilibrium flow  with  STM excitation parameter $f=8.4$. 
Velocity vectors are shown  with contours of  streamwise velocity. 
The eddy field is a sum of the EOF's with   amplitudes commensurate to their contribution to the total maintained eddy energy.
  The eddy field is dominated by the mean flow and the meandering of the streak  in this laminar state is very slight. 
Parameters are as in  Fig. \ref{fig:15}.} 
\label{fig:21}
\end{figure*}

Finally we note that  when the total field of the equilibrium mean flow and a typical realization of the eddy field are
plotted together only a weak undulation of the streak structure can be discerned as is observed with laminar streaks before transition.
For example a sample realization of the total equilibrium flow at STM excitation parameter $f=8.4$ is shown in Fig. \ref{fig:21}.

\section{Mechanism of roll/streak  equilibration}

We wish to gain an understanding of the dynamics underlying equilibration of the structural instability 
of the spanwise independent flow in part to provide insight 
into turbulent equilibria in general and in part as a first step in understanding how loss of structural stability by these roll/streak equilibria at $f=8.45$
leads to transition to to a  time dependent state.   

\begin{figure*}
 
\includegraphics[width=6in]{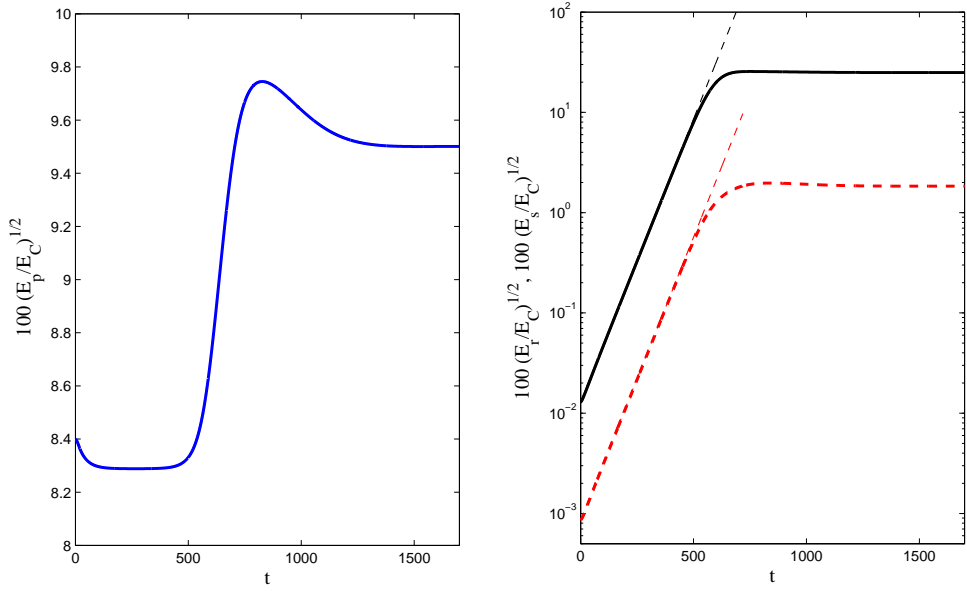}
\vspace*{-1mm} 
\caption{ 
(Color online) Left: Evolution of percent RMS eddy kinetic energy $100 \sqrt{E_p/E_C}$  normalized by the RMS kinetic energy of the Couette flow
after the structurally unstable spanwise independent equilibrium is 
perturbed by the most unstable streak perturbation shown in Fig. \ref{fig:8}.   The flow equilibrates to the roll/streak equilibrium 
shown in Fig. \ref{fig:9}c and Fig. \ref{fig:13}. Right:  Evolution of  the RMS  streak energy $100 \sqrt{E_s/E_C}$ and RMS roll energy $10^2 \sqrt{E_r/E_C}$ 
is exponential with growth rate $\lambda=0.014$. Parameters are as in Fig. \ref{fig:8},  the STM excitation parameter is $f=8.4$..} 
\label{fig:22}
\end{figure*}

\begin{figure*}
 
\includegraphics[width=6in]{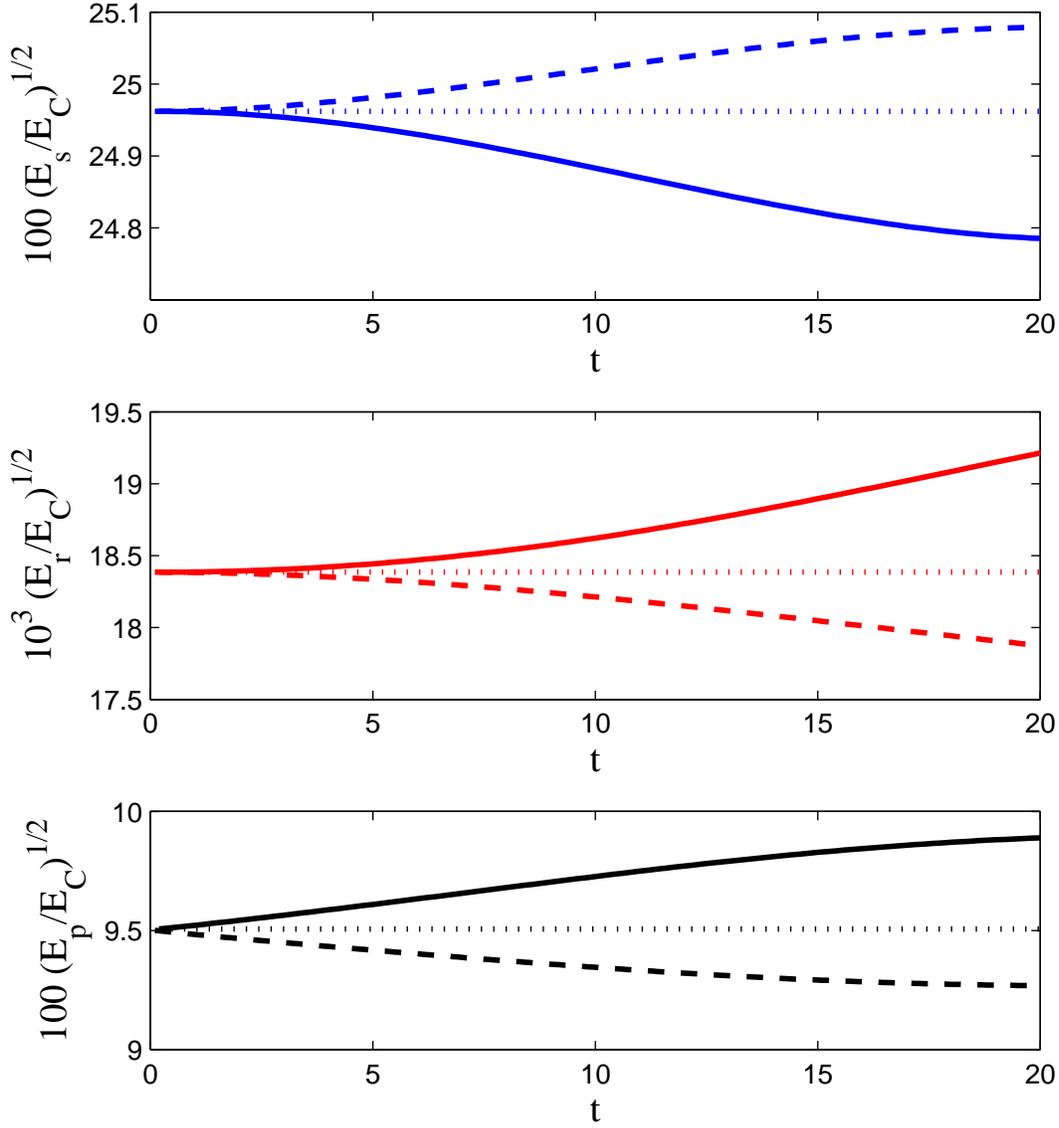}
\vspace*{-1mm} 
\caption{ 
(Color online) Percent RMS  normalized streak amplitude $100 \sqrt{E_s/E_C}$ (top panel),
permil RMS normalized  roll amplitude $10^3 \sqrt{E_r/E_C}$ (middle panel), and percent RMS normalized
perturbation velocity amplitude $100 \sqrt{E_p/E_C}$ (bottom panel)
as a function of time for decrease (solid) and increase (dashed) in the damping rate of the inflectional mode compared to its  damping rate at equilibrium.  The corresponding energies  at equilibrium are also shown (dotted).  The inflectional mode clearly damps rather than drives the streak.
Parameters as in Fig. \ref{fig:15}.
} 
\label{fig:23}
\end{figure*}

We will describe the equilibration of the structural instability of the spanwise independent flow at STM excitation parameter $f=8.4$ and in particular show how
the inflectional mode is instrumental in producing the equilibrium.  When the most unstable SSST eigenfunction 
is introduced into the spanwise independent flow, it grows exponentially  at rate $\lambda=0.014$ as predicted by SSST theory and finally equilibrates as can be seen in  Fig. \ref{fig:22}.   As the STM excitation parameter approaches $f=8.4$  oblique non-normal wave perturbations dominate the structures forcing the roll through their Reynolds stress 
and these produce a roll/streak equilibrium with  highly inflected streaks (cf Fig. \ref{fig:9}d).  
In association with  this inflection the primary inflectional mode (with structure as in Fig. \ref{fig:16}) 
approaches the stability boundary (cf Fig. \ref{fig:10}d).   
Because it is drawing energy from the streak this mode produces strong  downgradient 
Reynolds stresses that  damp the mode while the Reynolds stress it produces to force the roll circulation are relatively weak.  In order to compare the relative contribution of the direct downgradient Reynolds stress due to the inflectional mode in damping the streak with the mode's indirect Reynolds stress effect  in forcing the roll circulation and thereby building  the streak we artificially impose  a modification of the real part of the eigenvalue of the mode at equilibrium, $\sigma_{rE}$,
specifically we set the growth of this mode equal to  $9/10 \sigma_{rE}$
and $11/10 \sigma_{rE}$,
and integrated forward the  SSST equations in order to determine the mean flow tendency.  When the mode is  less stable the
perturbation energy increases, the associated roll circulation also increases as it is directly forced by the oblique structure of the 
perturbations, but the streak decreases because the enhanced downgradient fluxes by the less damped mode dominate over the increase in the streak induced by the roll circulation.
The opposite happens when the mode is made more stable,  as shown in Fig. \ref{fig:23}.

We conclude that  the primary mechanism of streak stabilization at high FST is the
inflectional mode.  
As the $f$ rises above $f = 8.45$ the inflectional mode is no longer able to 
stabilize the streak and a second structural instability ensues at $f_u=8.45$ in which 
the oblique waves further accelerate the roll/streak complex.  
However, the streak remains perturbation stable until a very high amplitude is reached at which point the flow becomes 
time dependent and aperiodic so that the notion of an unstable temporal mode is no longer 
well defined.  We will not examine this time dependent regime further in this work.

\section{Mechanism producing the spanwise streak spacing}

A fully turbulent boundary layer, such as that approximated by the Reynolds-Tiederman profile, is maintained  by the ensemble of  eddies in the boundary layer. 
It is commonly observed in turbulent boundary layers that spanwise streak spacing is approximately 100 $y^+ $ with wall unit $y^+ \equiv  \sqrt{{\nu}/({{\partial U}/{\partial y}})}$ and 
${\partial U}/{\partial y}$
being evaluated at the boundary.    As the boundary layer is itself approximately 50 $y^+$ in wall normal extent this  spacing is consistent with a roll of unit aspect ratio confined to the boundary layer\citep{Kim-etal-1987,Hutchins_Marusic-2007}.  

\begin{figure*}
 
\includegraphics[width=6in]{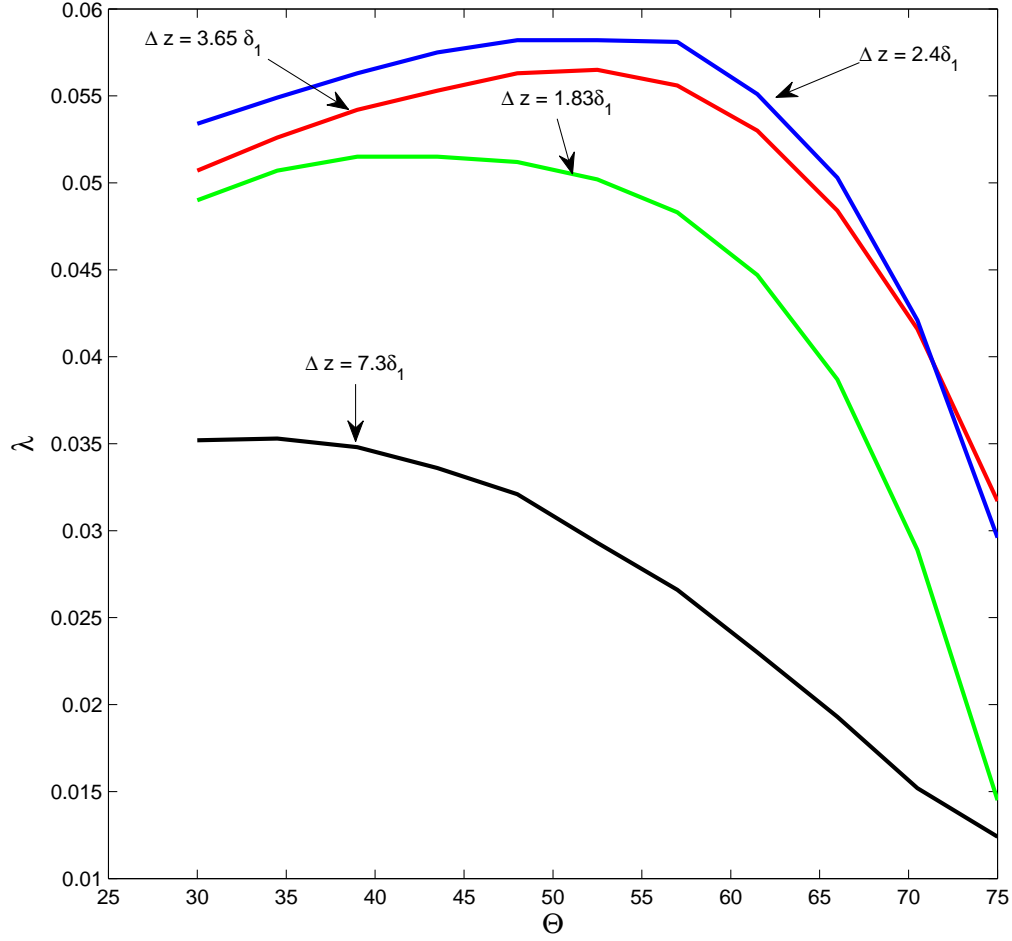}
\vspace*{-1mm} 
\caption{ 
(Color online) Growth rate, $\lambda$, of the structurally unstable streaks in a Blasius boundary layer  as a function
of perturbation structure obliqueness,  $\Theta\equiv\tan^{-1}(m/k)$.    
The displacement thickness is $\delta_1=1.72$ and $m$  is  the wavenumber of the streak.
Maximum  growth rate occurs for $m=3$ which corresponds to streak spacing  $\Delta z=2.4 \delta_1$ or $50 y^+$ wall units consistent with observations. 
Also shown are the growth rates for $m=1,2,4$.
 The channel width is $L_z=4 \pi$, the Reynolds number is $R=400$,  the STM excitation parameter is
$f=10$ and the channel height is $L_y=7$.
} 
\label{fig:24}
\end{figure*}

We have concentrated on the formation of streaks from FST in which the deviation of the mean boundary layer flow profile from the stationary Couette flow is small compared with that found in fully turbulent boundary layers.  In order to study streak spacing  in a numerically resolved example we maintain a Blasius profile stationary with an appropriate body force\citep{Westin-etal-1994}, subject it to a supercritical STM excitation parameter,  and obtain the maximum growth rate of the structural instability as a function of spanwise wavenumber  $m$  of the unstable streak using the power method. 
 Growth rate of the most unstable SSST roll/streak eigenmodes at STM excitation parameter $f=10$ in a Blasius boundary layer at  Reynolds number 
$R_x=1.6 \times 10^5$  (based on the distance from the leading edge)   are shown in Fig. \ref{fig:24}.  
The maximum SSST instability occurs at the wavenumber corresponding to a spacing between  low speed streaks of
$\Delta z = 2.4 \delta_1$ where $\delta_1=1.72 \sqrt{ \nu x / U}$ is the displacement thickness,
consistent with unit roll aspect ratio.  Although the implied selectivity is not very strong this result also agrees with observations\citep{Westin-etal-1994}.
The maximum is 
achieved by oblique waves with obliqueness parameter $\Theta =  \tan^{-1} m / k$,  close to the value $53^0$ obtained from the simple argument of section I.  This agreement 
shows that the basic dynamics are captured  by the oblique plane wave solutions on an unbounded constant shear flow\citep{Kelvin-1887b, Moffatt-67, Farrell-Ioannou-1993a,Farrell-Ioannou-1993b}.  These structures are scale 
independent and the streaks  that are formed by the SSST instability share the universal character of these
oblique perturbations.

\section{Discussion}

There are a number of points we wish to emphasize in connection with the above results:

\begin{enumerate}

\item Streaks can arise from a spontaneous cooperative exponential instability of the roll/streak/shear plus turbulence complex.

\item At finite amplitude the streaks of the roll/streak/shear plus turbulence instability complex typically form stable nonlinear equilibria. 

\item The cross-stream/spanwise torques supporting the equilibrium roll/streak structure are  produced by oblique waves properly collocated with the roll/streak structure.

\item The structural stability boundary for streak formation by this cooperative instability is not associated with modal instability of the streak and the finite amplitude equilibrium roll/streak structure is stable despite being highly inflectional in the spanwise direction.    In fact, the inflectional mode acts to stabilize the streak.

\end{enumerate}

A question often raised is whether the mechanism forming the roll/streak complex is essentially linear non-normal or essentially 
nonlinear\citep{Waleffe-1995,Henningson-1996b}.
We find that both are involved in agreement with Kim and Lim\citep{Kim-Lim-2000}:  the roll/streak  is the optimal non-normal growth structure the growth of which is essentially related to non-normality of the linear shear flow dynamics while the destabilization  of  this  roll/streak structure  arises from the quadratically nonlinear Reynolds stresses.  This forcing by perturbation Reynolds stresses of the non-normal optimal roll structure is an example of  the mechanism of structure maintenance in turbulence by nonlinear scattering of perturbation energy into the linear non-normal optimally growing subspace.
Because the non-normal roll/streak structure is so highly amplified, the energy scattered by nonlinear interaction into this  structure dominates the structure of  boundary layer turbulence.

Hamilton et al.\citep{Hamilton-etal-1995} (HKW) obtain in an integration of a spatially constrained shear flow model what they identify as a regeneration mechanism for maintaining at finite amplitude the roll/streak complex.  Their mechanism consists of a dynamic process of continual streak growth and decay associated  \citep{Waleffe-Kim-1997} with forcing of the roll circulation by an unstable mode arising from spanwise inflectional streak instability.  Their work addresses the problem of streak self-maintenance and presupposes that a streak of finite amplitude sufficient to produce an inflectional instability already exists.   This work addresses the formation of streaks from arbitrarily small initial perturbations in FST.   The finite amplitude equilibria we find, while inflectional, are modally stable and our streaks are maintained by the wave/mean flow interaction arising from a large subset of non-normal transient perturbation structures rather than by a single inflectional mode.   In fact, as we have seen, the inflectional mode is primarily responsible for stabilizing the SSST streak instability at finite amplitude. 
Nevertheless, their numerical simulations agree with both observations and SSST in a number of other particulars including the importance of the streamwise roll as the linear non-normal optimally growing structure necessary for the formation of streaks and of Reynolds stresses in forcing the streamwise roll.   HKW
find it remarkable that turbulent perturbations ``..produce additional streamwise vorticity in exactly
the right places to augment the streamwise vortices."\citep{Hamilton-etal-1995} but as we have seen, at least in the initial stages of streak formation,  this coincidence is a necessary consequence of the existence of the SSST eigenmode and its nonlinear extension.  

The SSST system provides analysis tools for obtaining a fundamental understanding of the underlying mechanisms of turbulence that can not be gained from interpretation of simulations alone so that we may say that SSST constitutes a theory of turbulence as distinct from a description of turbulence or an interpretation in dynamical terms of observations of turbulence.  In this work we have used SSST to study the formation and maintenance of roll/streak structures in FST which is important  in its own right but also as a component of the mechanism of bypass transition which typically proceeds from a pre-existing roll/streak structure.  

 \section{Conclusion}

Emergence of the roll/streak coherent structure in turbulent flow  is a problem of great theoretical and practical importance.  In this work we applied SSST to demonstrate a mechanism by which  the roll/streak structure can arise from an emergent instability of roll/streak/turbulence interaction in boundary layer shear flow.   


This emergent SSST instability giving rise to the roll/streak structure exploits the optimality of the non-normal roll/streak structure growth mechanism not by introducing an individual chance perturbation in cross-stream velocity but rather by organizing the ubiquitous torques associated with turbulent Reynolds stress divergence in the cross stream/spanwise plane to produce and maintain the optimal roll structure.  This organization of the Reynolds stress by the streak resulting in forcing of the roll provides the missing coupling between the streak and roll that is required to produce instability from the non-normal growth process.  
At small amplitude the roll/streak structure  grows exponentially as an eigenmode but at finite amplitude this growth is arrested and the structure approaches a nonlinear equilibrium.  This mechanism of streak formation and equilibration is not related to instability of the perturbation dynamics of the streak and this cooperative SSST instability occurs in the absence of shear flow instability and the inflectional mode is found to stabilize the streak.  
%

\begin{acknowledgments}
\noindent This work was supported  by NSF ATM-0123389. 
\end{acknowledgments}

\clearpage

\oddsidemargin-0.5in

\oddsidemargin-0.5in

\oddsidemargin-0.5in

\clearpage

\clearpage

\clearpage

\end{document}